\documentclass[draft]{agujournal2019}
\usepackage{url} 
\usepackage{lineno}
\usepackage[inline]{trackchanges} 
\usepackage{soul}
\addeditor{AB}
\soulregister\cite7
\soulregister\citeA7
\soulregister\ref7
\linenumbers

 \usepackage{lmodern}
\usepackage{amssymb, amsmath}
\usepackage[bb=dsserif]{mathalpha}
\usepackage{bbold}
\usepackage{bm}
\usepackage{xr}

\makeatletter
\newcommand*{\addFileDependency}[1]{
\typeout{(#1)}
\@addtofilelist{#1}
\IfFileExists{#1}{}{\typeout{No file #1.}}
}
\makeatother

\newcommand*{\myexternaldocument}[1]{%
\externaldocument{#1}%
\addFileDependency{#1.tex}%
\addFileDependency{#1.aux}%
}

\myexternaldocument{si_template_2019}

\usepackage{pdfpages} 
\usepackage{pgffor} 

\makeatletter
\AtBeginDocument{\let\LS@rot\@undefined}
\makeatother

\def\supplementfilename{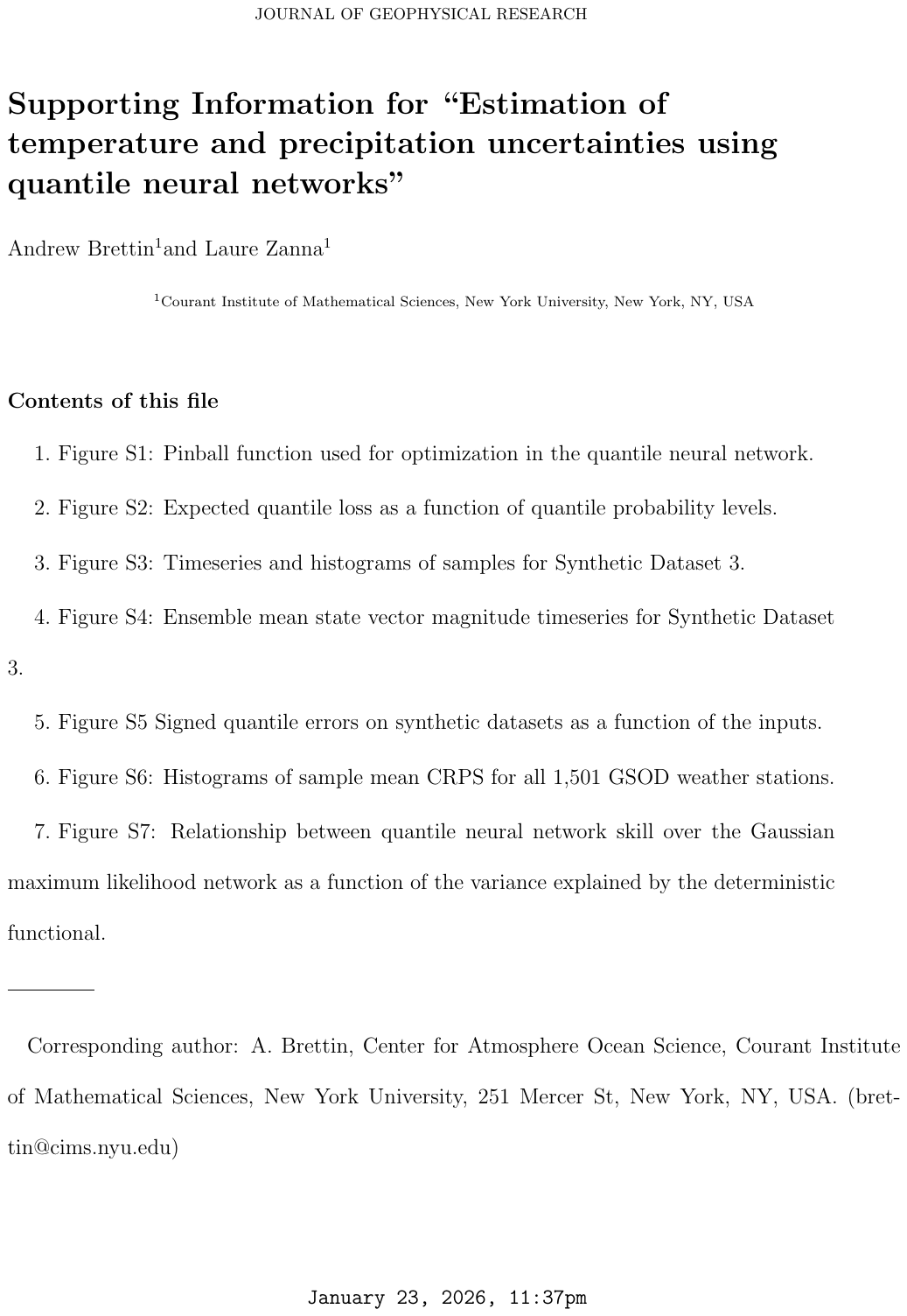}

\pdfximage{\supplementfilename}
\def\numbersupplementpages{\the\pdflastximagepages}

\newif\ifarXiv
\arXivtrue

\draftfalse

\journalname{JGR: Machine Learning and Computation}

\begin{document}

\nolinenumbers

\title{Estimation of temperature and precipitation uncertainties using quantile neural networks}

\authors{Andrew E. Brettin\affil{1}, Laure Zanna\affil{1, 2}}

\affiliation{1}{Courant Institute of Mathematical Sciences, New York University, New York, NY, USA}
\affiliation{2}{Center for Data Science, New York University, New York, NY, USA}

\correspondingauthor{Andrew Brettin}{brettin@cims.nyu.edu}



\begin{keypoints}
      \item We evaluate a proposed ReLU-bias loss quantile neural network (RBLQNN) for constraining uncertainties on synthetic and observational data
      \item The RBLQNN's ease of implementation and generality reveal its suitability for quantifying uncertainties
      \item The RBLQNN reveals nonlinear temperature dependencies on observed quantities, and nonlinear, non-Gaussian dependencies for precipitation
\end{keypoints}

%
%

%
%

\begin{abstract}
Extreme events pose significant risks and are challenging to predict. Assessing climate hazards requires placing quantitative constraints on geophysical fields under observable but fluctuating conditions. We propose a framework for estimating uncertainties—a ReLU-bias loss quantile neural network (RBLQNN)—with two novel modifications to the loss function to enforce uniform quantile accuracy and reduce degenerate predicted probability distributions. We evaluate the RBLQNN against other probabilistic baselines on a suite of datasets: synthetic datasets, observed daily temperature maxima from 1,501 NOAA Global Surface Summary of the Day (GSOD) weather stations, and altimetry-observed precipitation from the Tropical Rainfall Measuring Mission (TRMM). On synthetic datasets, the RBLQNN accurately predicts conditional distributions where more restrictive methods like linear quantile regression (LQR) or mean-variance estimation (MVE) neural networks fail, mitigates shortcomings of some other quantile neural networks, and converges stably under a range of hyperparameters. When applied to daily temperature maxima, the RBLQNN reveals that temperature distributions are relatively well described by Gaussian statistics, though nonlinear dependencies on local sea level pressure and geopotential heights appear important. For precipitation statistics, the RBLQNN strongly outperforms both LQR and MVE baselines, demonstrating its capacity to capture highly nonlinear and non-Gaussian conditional distributions. The RBLQNN's performance across varied datasets demonstrates it is a flexible and general approach for constraining uncertainties in geophysical quantities with nonlinear or non-Gaussian conditional dependencies.
\end{abstract}

\section*{Plain Language Summary}
The climate system is highly chaotic and unpredictable, often yielding extremes and risks that must be quantified. In light of these uncertainties, we propose a data-driven probabilistic technique, a type of ``quantile neural network," for quantifying uncertainties that requires few assumptions and has a straightforward implementation. Using a synthetic dataset, we demonstrate the advantages of this quantile neural network over baselines that make stronger assumptions, such as linear relationships between inputs and outputs and normally distributed uncertainties. We then apply this technique to weather station temperature data and satellite observations of precipitation, finding that daily maximum temperatures are well described by nonlinear relationships with normally distributed uncertainties, whereas precipitation depends significantly nonlinearly on the model inputs and exhibits non-normal statistics. This work shows how quantile neural networks can be easily implemented to gain a more accurate representation of uncertainties in the geosciences.

%
%

\section{Introduction}
\label{sec:introduction}

The climate system is governed by complex, highly nonlinear interactions between the atmosphere, ocean, land, and cryosphere \cite{gupta2022perspectives}, and the chaotic dynamics that result can be difficult to predict with certainty given limited information about the system state \cite{vitart2017subseasonal}. Interactions between processes occur over a range of temporal and spatial scales, producing variability and extremes, often with adverse impacts to human populations \cite{newman2023global}. Given the lack of predictability in the climate system, accurately quantifying the uncertainty of geophysical fields under changing measurable conditions is crucial due to the hazards that can result from its chaotic dynamics.

Often, uncertainties in the atmosphere-ocean system are represented using Gaussian statistics. Gaussian assumptions underlie climate projections \cite{kopp2014probabilistic,ipcc2022spm}, stochastic parameterization of sub-gridscale processes \cite{franzke2015stochastic}, measurement approximations in data assimilation models and reanalysis products \cite{bocquet2010beyond,hersbach2020era5}, linear inverse models \cite{penland1989random, penland1995optimal}, in-situ and altimetric observational product quality control and more. The Gaussian distribution's relevance to characterizing a wide range of uncertainties in the Earth system stems from a few theoretical considerations \cite{sura2015perspectives}. Firstly, the normal distribution plays a crucial role in the Central Limit Theorem, which states that the sample mean of independently and identically distributed random variables approaches a normal distribution as the sample size increases. The Central Limit Theorem implies that aggregation operations (e.g., the averaging involved in the measurement, simulation, and forecasting of geophysical variables) tend to produce normally distributed quantities \cite{delsole2022statistical1}. Another reason that the Gaussian distribution arises is due to its unique role as the maximum entropy distribution for a given mean and variance for quantities with unbounded support \cite{sura2015perspectives, majda2006nonlinear}. The principle of maximum entropy, first proposed by \citeA{jaynes1957information}, is that the maximum entropy distribution (i.e., the least informative one under given constraints) is the most probable distribution. Thus, without further information constraining the form of the uncertainties, the Gaussian is the ``best guess" of the underlying uncertainty.

One approach to quantifying Gaussian uncertainties that has gained significant traction for geoscience applications in recent years is mean-variance estimation (MVE) neural networks. Proposed by \citeA{nix1994estimating}, MVE networks are data-driven models optimized over the Gaussian negative log-likelihood to yield not only a point estimate, but also a standard deviation to quantify the level of uncertainty given the inputs. MVE networks have been used for developing stochastic parameterizations \cite{guillaumin2021stochastic,perezhogin2023generative,wu2025data}, identifying drivers of predictability \cite{gordon2022incorporating}, identifying exceedance times of critical global warming thresholds \cite{diffenbaugh2023data} and more \cite{haynes2023creating,barnes2021controlled,schreck2024evidential}.

\begin{figure}[!ht]
    \centering
    \includegraphics[width=\linewidth]{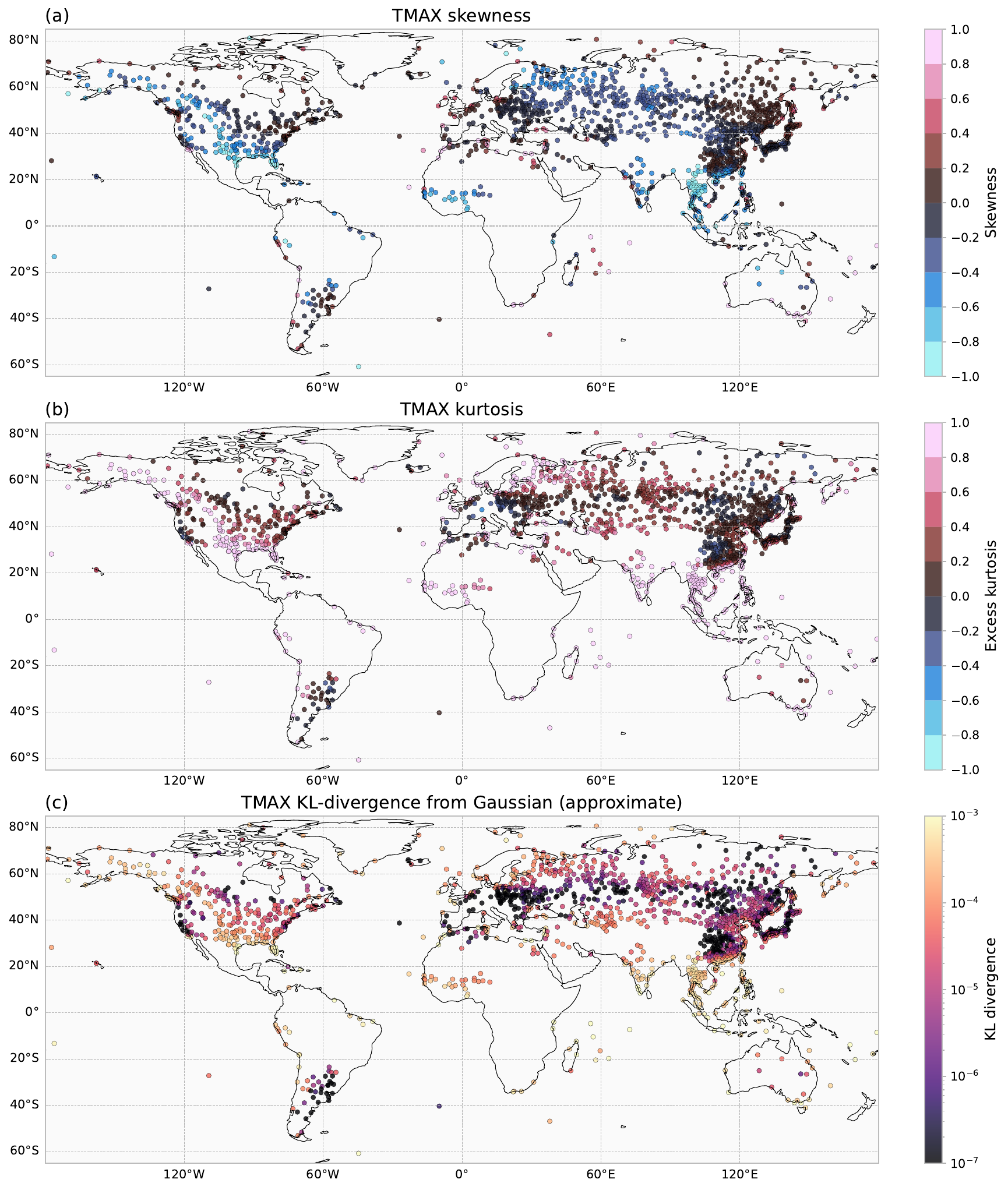}
    \caption{Non-Gaussianity of daily maximum surface air temperature (TMAX) seasonal anomalies, as measured by the sample's (a) skewness, (b) kurtosis, and (c) Kullback-Leibler divergence between the sample and a Gaussian estimated with the approach of \citeA{hyvarinen2000independent} using contrast function $G(u) = \log \cosh{u}$ (higher values indicate greater deviations from Gaussianity).}
    \label{fig:gsod_skewness_kurtosis}
\end{figure}

Despite the proliferation of machine learning methods that assume Gaussianity, however, many geophysical quantities are non-Gaussian. Observations of surface air temperature exhibit significantly non-Gaussian characteristics (Fig.~\ref{fig:gsod_skewness_kurtosis}; \citeA{proistosescu2016identification,catalano2021diagnosing,cavanaugh2014northern,mckinnon2016changing}), with numerous physical causes (such as tracer advection-diffusion processes and jet dynamics) supported by numerical simulations \cite{linz2018large,garfinkel2017non, hassanzadeh2015blocking} as well as theoretical arguments \cite{sura2015perspectives, kimura1993statistics,mclaughlin1996explicit,hu2002advection}. Precipitation statistics are highly non-Gaussian, and modeling precipitation statistics remains an active area of research \cite{ashkenazy2024data,li2023determining,scheuerer2020using,beck2020ppdist,martinez2019precipitation}. Deviations from Gaussianity have implications for the quantification of extremes \cite{bjarke2023record,loikith2019non}, as well as the changes in likelihood of tail events under changing temperatures \cite{loikith2015short}.

Studies such as \citeA{barnes2023sinh} have relaxed the Gaussian assumption of MVE networks by training neural networks to estimate parameters of probability distributions which include skewness and kurtosis parameters (e.g., the ``SHASH" distribution; \citeA{jones2019sinh}). Such approaches generalize the types of aleatoric uncertainties that can be estimated; nevertheless, they require parametric assumptions about the underlying uncertainties which may not necessarily hold. For instance, the SHASH distribution may inadequately represent precipitation statistics, particularly because nonnegative precipitation measurements are incompatible with its unbounded support. Alternatively, quantile regression techniques allow estimation of the response distribution without parametric assumptions about the distribution of uncertainties by using an optimization formulation for the quantile \cite{koenker1978regression,koenker2005quantile}. Linear quantile regression has been used for identifying temporal changes in distributions of surface air temperature observations \cite{mckinnon2016changing}, and for sea surface heights simulated by climate models \cite{falasca2023exploring}. The quantile regression loss has also been used to train regression neural networks, starting with \citeA{white1992nonparametric} and \citeA{taylor2000quantile}. Such quantile regression neural networks give an appealing means of estimating uncertainties, as they allow for complete estimation of nonlinear functional dependencies and non-Gaussian statistics. Several studies have used the quantile regression loss for neural networks to predict conditional probabilities, including for assessing financial risk \cite{chronopoulos2024forecasting,keilbar2022modelling}, energy demands \cite{zhang2019improved,tambwekar2021estimation,belloni2019conditional}, and healthcare outcomes risks \cite{zhang2025monotone,corsaro2024quantile}. Nevertheless, the use of quantile regression neural networks for geoscience applications seems to be limited to a few studies \cite{cannon2018non,haynes2023creating,bremnes2020ensemble, papacharalampous2025ensemble}. Technical challenges, such as limited sample size or ensuring quantile monotonicity \cite{chernozhukov2010quantile,cannon2018non,padilla2022quantile} may inhibit broader usage of quantile regression neural networks in the geosciences.

In this paper, we propose a flexible and simple quantile regression neural network for estimating uncertainties. Our implementation of the quantile regression neural network uses the Rectified Linear Unit (ReLU) as a loss function to encourage quantile monotonicity during training, and we refer to it as the ``ReLU bias loss quantile neural network" (RBLQNN). We evaluate the RBLQNN against baselines which assume Gaussian conditional distributions or linear dependence to assess the relative importance of linearity or Gaussianity assumptions in a variety of synthetic and observational datasets. After establishing the advantages of the RBLQNN on synthetic datasets, we assess the Gaussianity and linearity assumptions of two observational datasets: NOAA Global Surface Summary of the Day (GSOD) daily temperature measurements at 1,501 weather stations, and Tropical Rainfall Measuring Mission (TRMM) precipitation altimeter observations.

In Section~\ref{sec:methods}, we formulate our approach to conditional probability estimation, and describe the RBLQNN, baselines, datasets, and metrics. Then, in Section~\ref{sec:synthetic_results} we evaluate the performance of the RBLQNN, demonstrating its advantages over the baselines and other quantile regression neural network techniques. In Section~\ref{sec:observational_results}, we examine the importance of Gaussianity and linearity assumptions in the GSOD and TRMM datasets. We end with a discussion that provides some perspectives on our results and describes the caveats of the RBLQNN in Section~\ref{sec:discussion}.

\section{Methods}
\label{sec:methods}

\subsection{Conditional probability estimation}
\label{subsec:methods.conditional_probability}

\subsubsection{Formulation and optimization setup}
\label{subsubsec:methods.conditional_probability.formulation}
In this framework, geophysical target variables $Y$, such as surface air temperature or precipitation, are considered a function $r$ of random variables
\begin{equation}
    Y = r(\mathbf{X}, \Psi),
\end{equation}
where the $\mathbf{X} = (X_1, \dots, X_p)$ represent observable random variates and $\Psi$ represents the remaining aleatoric uncertainty. We seek to represent the distribution of the prediction variables conditioned on observed quantities $Y | \mathbf{X}=\mathbf{x}$.

We express probabilities in terms of quantiles. For a continuous cumulative distribution function (CDF) $F: \mathbb{R} \to (0, 1)$, the $q$-quantile $y_q$ is defined by $y_q = F^{-1}(q)$. Because the CDF is unique for a given distribution, a probability distribution is fully characterized by the set of quantiles for $q \in (0, 1)$. Thus, the conditional distribution of $Y|\mathbf{X}$ can then be formulated in terms of its quantiles $(Y|\mathbf{X})_q$ by functions $f^{(q)}: \mathbb{R}^p \to \mathbb{R}$ as 
\begin{equation}
    y_q = (Y|\mathbf{X}=\mathbf{x})_q = f^{(q)}(\mathbf{x}).
    \label{eq:conditional_quantiles}
\end{equation}
Identifying the conditional distribution amounts to determining $f^{(q)}$ for $q \in (0, 1)$.

The key insight behind quantile regression techniques is that the $q$-quantile $y_q$ satisfies the following optimization problem \cite{koenker1978regression}:
\begin{equation}
y_q = F^{-1}(q) \iff y_q = \underset{u}{\mathrm{argmin}} \ \mathbb{E}[\rho_q(Y - u)] = \underset{u}{\mathrm{argmin}} \ \int_\mathbb{R} \rho_q(y - u) \ dF(y),
    \label{eq:optimization_quantile}
\end{equation}
where the ``pinball function'' $\rho_q(t)$ is defined by 
\begin{equation}
\rho_q(t) = qt\mathbb{1}_{\{t \geq 0\}} - (1 - q)t\mathbb{1}_{\{t < 0\}}.
\label{eq:pinball_function}
\end{equation}
The pinball function is plotted for various values of $q$ in Supporting Figure \ref{supp:fig:pinball_function}.
Intuitively, the pinball function asymmetrically penalizes data above and below a given value to obtain estimates for a specified quantile. For instance, when predicting a high quantile like $q=0.9$, data $y$ is heavily penalized for exceeding the minimization argument $u$ in \eqref{eq:optimization_quantile}, but lightly penalized for subordinating $u$. This pushes the optimal value of $u$ towards the higher end of the data. When $q=0.5$, the pinball function is symmetric and reduces to the absolute value of its argument (scaled by a factor of 0.5).

This optimization formulation \eqref{eq:optimization_quantile} is analogous to how the mean of a distribution is the argument minimizer of the variance functional:
\begin{equation}
    \mu = \mathbb{E}Y \iff \mu = \underset{u}{\mathrm{argmin}} \ \mathbb{E}[(Y - u)^2] = \underset{u}{\mathrm{argmin}} \ \int_\mathbb{R} (y - u)^2 \ dF(y). 
    \label{eq:optimization_mean}
\end{equation}

The optimization formulation \eqref{eq:optimization_quantile} provides an avenue for determining conditional distributions using regression techniques. In a general regression problem, we seek to estimate some property of the conditional distribution of $Y|\mathbf{X}$ by a functional $f(\mathbf{X})$. This minimization formulation allows this functional to be empirically optimized over a class of candidate functions $\{f_{\mathbf{\theta}}\}_{\mathbf{\theta} \in {\Theta}}$ for a given dataset $\{(\mathbf{x}_i, y_i)\}_{i=1}^n$ sampled from the joint distribution $(\mathbf{X},Y)$. For example, in least-squares regression, the optimization formulation for the mean in \eqref{eq:optimization_mean} is used to empirically estimate the conditional mean:

\begin{equation}
\mathbb{E}[Y|\mathbf{X} =\mathbf{x}]  = f(\mathbf{x}) \approx \hat{f}_\mathbf{\theta}(x),
\end{equation}
where
\begin{equation}
\hat{f}_\mathbf{\theta} = \underset{f_\mathbf{\theta}}{\mathrm{argmin}} \left\{ \frac{1}{n}\sum_{i=1}^n \left(y_i - f_{\mathbf{\theta}}(\mathbf{x}_i)\right)^2 \right\}.
\label{eq:quantile_erm}
\end{equation}
In a similar vein, the optimization formulation for the quantiles \eqref{eq:optimization_quantile} can be used to estimate the conditional quantile of the distribution:
\begin{equation}
    y_q = (Y|\mathbf{X}=\mathbf{x})_q = f^{(q)}(\mathbf{x}) \approx \hat{f}_{\mathbf{\theta}}^{(q)}(\mathbf{x}),
\end{equation}
where
\begin{equation}
\hat{f}_\mathbf{\theta}^{(q)} = \underset{f_\mathbf{\theta}}{\mathrm{argmin}} \left\{ \frac{1}{n}\sum_{i=1}^n \rho_q \left(y_i - f_{\mathbf{\theta}}(\mathbf{x}_i)\right) \right\}.
\end{equation}

\subsubsection{ReLU bias loss quantile regression neural network (RBLQNN)}
\label{subsubsec:methods.conditional_probability.qrnn}
The RBLQNN is a multilayer perceptron $f_\theta : \mathbb{R}^p \to \mathbb{R}^m$ that outputs predicted quantiles for a discrete set of probability levels $0 < q_1 < \dots < q_m < 1$:
\begin{equation}
	(\hat{y}_{q_1}, \dots, \hat{y}_{q_m}) = f_\theta(\mathbf{x}).
\end{equation}
Quantiles at intermediate probability levels $q\neq q_j$ can then be approximated using interpolation techniques. In this study, we evaluate predictions of the quantiles using $m=19$ equispaced probability levels with increments of $\Delta q = 0.05$. This is a subjective choice loosely informed by statistical conventions \cite{fisher1970statistical, xu2017composite}.

Since the neural network predicts multiple quantiles simultaneously, the loss function optimizes predictions for all quantiles:
\begin{equation}
    \mathcal{L}_Q(y_i, \boldsymbol{\hat{y}}_{q,i}) = \sum_{j=1}^m \lambda_j \rho_{q_{j}}(y_i - \hat{y}_{q_j, i}).
    \label{eq:composite_quantile_loss}
\end{equation}

We propose to combine two simple additions for mitigating issues that may arise when using this loss for training:
\begin{enumerate}
    \item \textit{Quantile counterbalancing}. At the quantile $y_q = F^{-1}(q)$, the expected value of the quantile loss $\mathbb{E}\left[ \rho_q(Y - y_q)\right]$ is different for different probability levels $q$. In order to ensure that each quantile is optimized evenly, the loss weights $\lambda_j$ in \eqref{eq:composite_quantile_loss} should be chosen to be inversely proportional to the expectation $\lambda_j = \mathbb{E}\left[ \rho_q(Y - y_q)\right] ^{-1}$. This expectation reduces to 
    \begin{equation}
        \mathbb{E}\left[ \rho_q(Y - y_q)\right] = q \left(\mathbb{E}Y - \mathbb{E}[Y|Y < y_q] \right).
    \end{equation}
    While the true expectation depends on the distribution of $Y$ (which is unknown),  Supporting Figure~\ref{supp:fig:weight_approximation} shows that this expectation is not extremely sensitive to the type of distribution. For the purposes of developing bona fide weights, we assume a standard-normal distribution so that $\lambda_j = \exp{\left( \frac{\Phi^{-1}(q_j)}{2}\right)}$, where $\Phi$ is the CDF of the standard normal distribution.
    \item \textit{ReLU bias loss}. Because the CDF must be monotonic, probability distributions become degenerate when predicted quantiles cross. That is, for $q_1 < \dots < q_m$, we must have that $\hat{y}_{q_1} < \dots <\hat{y}_{q_m}$.  Various methods for avoiding quantile crossings have been proposed, such as monotone rearrangement of predicted quantiles \cite{chernozhukov2010quantile}, applying monotonicity constraints to function inputs \cite{cannon2018non} or predicting nonnegative increments for successive quantile values \cite{padilla2022quantile}. These methods range in complexity; here, we apply a simple bias loss to discourage quantile crossings during training while retaining the transparent output structure of the quantile regression neural network:
    \begin{equation}
        \mathcal{L}_\mathrm{ReLU}(\boldsymbol{\hat{y}}_{q,i}) = \sum_{j=1}^{m-1} \mathrm{ReLU}(\hat{y}_{q_j,i} - \hat{y}_{q_{j+1},i}),
        \label{eq:relu_bias_loss}
    \end{equation}
    where the Rectified Linear Unit function is given by $\mathrm{ReLU}(x) = x \boldsymbol{1}_{\{x \geq 0\}}$. If any of the quantiles is greater than the succeeding quantile, the ReLU loss is positive. If all quantiles are ordered, then the ReLU penalty is 0.
\end{enumerate}

The net loss function that is used for training the RBLQNN is given by 
\begin{equation}
    \mathcal{L}(y_i, \boldsymbol{\hat{y}}_{q,i}) = \mathcal{L}_Q(y_i, \boldsymbol{\hat{y}}_{q,i}) + \eta \mathcal{L}_\mathrm{ReLU}(\boldsymbol{\hat{y}}_{q,i}), 
    \label{eq:net_quantile_loss}
\end{equation}
where $\eta > 0$ is a hyperparameter.

\subsubsection{Baselines}
\label{subsubsec:methods.conditional_probability.baselines}
The predicted distributions made by the quantile neural network are compared against two baselines: linear quantile regression (LQR) and mean-variance estimation (MVE) neural networks. Linear quantile regression allows for arbitrary conditional probability estimates, but the functional dependence on the regressors is assumed to be linear. On the other hand, mean-variance estimation networks allow for nonlinear dependencies, but restrict conditional probabilities to be Gaussian. Together, these baselines help assess the relative importance of nonlinearities and non-Gaussianity in the predicted conditional distributions.

\paragraph{Linear quantile regression (LQR)}
For linear quantile regression, it is assumed that the quantiles of the regressand depend linearly on the regressors:
\begin{equation}
    y_q = (Y|\mathbf{X}=\mathbf{x})_q = \beta_0(q) + \mathbf{\beta}(q)^T \mathbf{x}.
\end{equation}
Accordingly, the family of functions minimized using \eqref{eq:quantile_erm} are limited to linear functions:
\begin{equation}
\hat{\beta}_0(q), \hat{\beta}_1(q) = \underset{\beta_0(q), \beta_1(q)}{\mathrm{argmin}} \frac{1}{n} \sum_{i=1}^n \rho_q(y_i - \hat{y}_{q,i}),
\end{equation}
where $\hat{y}_{q,i} = \beta_0(q) + \beta_1(q) x_i$.

For the linear quantile regression model, predictions are made for each quantile $q_1, \dots, q_m$ simultaneously, so that weights are shared for different quantile predictions, which has been shown to improve predictions \cite{zou2008composite, jiang2012oracle}.

\paragraph{Mean-variance estimation (MVE) neural networks}
As an alternative to linear quantile regression, we explore predictions made by mean-variance estimation (MVE) networks \cite{nix1994estimating}. MVE networks are artificial neural networks that yield predictions of conditional probability densities by outputting the parameters of a normal distribution $\hat{\mu}_i = \hat{\mu}(\boldsymbol{x}_i), \hat{\sigma}^2_i = \hat{\sigma}^2(\boldsymbol{x}_i)$. (In practice, because the predicted standard deviation should be positive, the log-variance $\log(\hat{\sigma}^2_i)$ is typically used as an output of the neural network and then exponentiated to yield positive values of $\hat{\sigma}_i^2$.) Weights and biases in the neural network are updated by optimizing the log-likelihood of the target data under the predicted parameters:
\begin{equation}
    \mathcal{L}\left(y_i, \hat{\mu}(\boldsymbol{x}_i), \hat{\sigma}^2(\boldsymbol{x}_i)\right) = - \log{p(y_i | \hat{\mu}_i, \hat{\sigma}^2_i)} = \frac{1}{2} \left[ \log(\hat{\sigma}_i^2) + \left(\frac{y_i - \hat{\mu}_i}{\hat{\sigma}_i}\right)^2\right] + C,
    \label{eq:gaussian_nll}
\end{equation}
where $p(y| \mu, \sigma^2) = \frac{1}{\sqrt{2\pi} \sigma} \exp \left[\frac{1}{2}\left(\frac{y - \mu}{\sigma}\right)^2\right]$ is the probability density function of a normal distribution with parameters $\mu$ and $\sigma^2$, and $C = \frac{1}{2}\log{2\pi}$ is an immaterial optimization constant. Quantile predictions are then given by $\hat{y}_q = \hat{\mu}_i + \hat{\sigma}_i\Phi^{-1}(q)$.

\paragraph{Alternative quantile regression neural network approaches}
We also assess the performance of our quantile regression neural network trained on \eqref{eq:net_quantile_loss} against various other quantile regression neural network techniques:
\begin{enumerate}
    \item \textit{Unweighted network}. This quantile neural network is trained using \eqref{eq:composite_quantile_loss} with uniform weighting $\lambda_1 = \dots = \lambda_m = 1$, and is termed the ``composite'' network in \citeA{xu2017composite}.
    \item \textit{No-bias network}. This network uses the normal-distribution inverse-expectation weighting scheme $\lambda_j = \exp{\left( \frac{\Phi^{-1}(q_j)}{2}\right)}$. However, the bias loss of \eqref{eq:relu_bias_loss} is not applied.
    \item \textit{Cumulative increment}. Proposed in \citeA{padilla2022quantile}, this network enforces strict monotonicity by predicting positive increments between successive quantiles. Specifically, the network outputs values $h_1(\mathbf{x}), \dots, h_m(\mathbf{x})$ such that 
    \begin{equation}
        \hat{y}_{q_j} = \begin{cases}
            h_1(\mathbf{x}) & (j=1), \\
            h_1(\mathbf{x}) + \sum_{k=2}^j \log(1 + e^{h_k(\mathbf{x})}) & (j=2, \dots, m).
        \end{cases}
        \label{eq:cuminc_output}
    \end{equation}
    Since $\log{(1+e^t)} > 0$, monotonicity of predicted quantiles is strictly enforced. This network can then be trained using the loss function in \eqref{eq:composite_quantile_loss} using equal weights $\lambda_1 = \dots = \lambda_m=1$.
\end{enumerate}

\paragraph{Mean squared error (MSE) neural networks}
In addition to the probabilistic baselines, a neural network trained using the Mean Squared Error (MSE) to predict the target variable is also used as a deterministic baseline.

\subsection{Datasets}
\label{subsec:methods.data}
We evaluate the methods for conditional probability estimation on a suite of different datasets: 1) synthetic datasets, 2) weather station daily temperature maxima, and 3) altimetry-observed precipitation.

\subsubsection{Synthetic datasets}
\label{subsubsec:methods.data.synthetic}
We first demonstrate the performance of the different methods on three different synthetic datasets:
\begin{itemize}
    \item \textit{Dataset 1}. $Y = X^2 + \Psi$, where $X \sim N(0, 1)$ is drawn from a standard normal and $\Psi$ is drawn from a Gumbel distribution. The Gumbel distribution, which is used in extreme value theory, was selected to represent the aleatoric uncertainty due to its non-Gaussianity and support on the entire real line. 
    \item \textit{Dataset 2}. $Y = \textrm{Beta}(\alpha, \beta)$, where $\alpha = X + 0.2$, $\beta=1.2 - X$, and $X \sim U(0,1)$ is uniform distributed on $[0,1]$. This dataset was formulated to investigate the impacts of heteroscedasticity on the conditional probability estimation techniques. The Beta distribution has bounded support, providing an interesting test case for the RBLQNN and other baselines.
    \item \textit{Dataset 3}. Often, uncertainties of interest arise from quantities generated by a time-varying dynamic process. Thus, we consider a stationary distribution generated from a two-dimensional stochastic gradient system of damped Langevin dynamics \cite{schlick2010molecular} of $\boldsymbol{x} = (x, y)$,
    \begin{equation}
        \frac{d\boldsymbol{x}}{dt} = - \nabla V(\boldsymbol{x}) + \xi,
        \label{eq:stochastic_gradient_system}
    \end{equation} 
    where $V(\boldsymbol{x})$ is a potential function and $\xi \sim N(0, 1)$ is white noise. The stationary joint distribution $f$ yielded by the dynamics is governed by the steady-state Fokker-Planck equation, with solution given by the Boltzmann distribution $f(x, y) \propto e^{-V(x, y)}$ \cite{landau2013statistical}. The conditional distributions can then be computed using numerical integration. We consider a potential of the form
    \begin{equation}
        V(x,y) = \prod_{i=1}^3\left((x - x_i)^2 + (y - y_i)^2\right)
        \label{eq:potential_function}
    \end{equation}
    which contains local minima at the points $(x_i, y_i)$. For this dataset, we set $(x_1, y_1) = (-\frac{1}{2}, -\frac{1}{2})$, $(x_2, y_2) = (1, -1)$, and $(x_3, y_3) = (1, \frac{1}{2})$, which yields a unimodal conditional distribution at $x = x_1$ and a bimodal conditional distribution at $x = x_2 = x_3$. We simulate the trajectories from the origin using an Euler-Maruyama solver with timesteps $\Delta t=10^{-4}$ up to a final time of $T=50,000$. The joint distribution and corresponding histogram with 10,000 bins (100 equispaced bins in $x$ and $y$) are shown in Supporting Figure~\ref{supp:fig:euler_maruyama}. Histogram frequencies are within 2\% of the true density for all bins, indicating that the samples are consistent with the theoretical distribution yielded by the Fokker-Planck equation. In order to verify that samples are drawn from the stationary distribution, Supporting Figure~\ref{supp:fig:steady_state} shows the ensemble mean and spread of the magnitude $\|(x,y)\|_2$ to show that initial condition information is lost after a few seconds of simulation time. The training, validation, and testing sets are randomly sampled from the values generated by this timeseries. 
\end{itemize}

Figures~\ref{fig:toy_data_quantile_predictions}a, \ref{fig:toy_data_quantile_predictions}e, and \ref{fig:toy_data_quantile_predictions}i show the quantiles of the conditional distribution of $Y|\mathbf{X}$ and samples drawn from the joint distribution $(\mathbf{X}, Y)$ for the three synthetic datasets. For each dataset, we examine performance using training sets of $n=10,000$ samples, validating training on a dataset of $1,000$ samples, and measuring performance on a test dataset of $1,000$ samples. The input and target samples of Datasets 1 and 3 are standardized using the training dataset mean and standard deviation. No standardization or normalization was applied to Dataset 2, as the samples already ranged from 0 to 1.

\subsubsection{NOAA Global Surface Summary of the Day daily temperature maxima}
\label{subsubsec:methods.data.gsod}
We next consider temperature observations from 1,501 weather stations over 1960--2020 given by the National Oceanic and Atmospheric Administration's (NOAA) Global Surface Summary of the Day (GSOD) dataset \cite{noaa1999gsod}. The GSOD dataset contains daily summary statistics for 18 surface meteorological variables derived from the NCEI Integrated Surface Hourly (ISH) dataset \cite{lott2001isd, smith2011integrated}. We use the daily maximum temperatures (TMAX) and daily-mean sea level pressures (SLP) from this dataset. In addition, we supplement the GSOD dataset with 500 mb and 850 mb geopotential heights (z500, z850) from ECMWF's ERA5 reanalysis \cite{hersbach2020era5}. These variables were included to provide synoptic information about the atmospheric circulation at the middle and lower troposphere. For each weather station, models are fitted to predict TMAX conditioned on concurrent station-derived SLP and geopotential heights from the nearest ERA5 gridpoint. Separate models are trained for each station.

The GSOD dataset contains records from more than 9,000 stations. We filter data-limited stations by requiring a minimum number of observations for training, validation, and testing. We use data from years 1960--2010 for training, 2011--2015 for validation, and 2016--2020 for testing, and stations are rejected if there are fewer than 30 years of daily records for training (10,950 samples) or 3 years for validation or testing (1,095 samples). This leaves 1,501 stations for our analysis. Inputs and targets are not detrended or deseasonalized, but values are standardized according to the mean and standard deviation of the training data for each station.

\subsubsection{TRMM altimetry precipitation observations with ERA5 reanalysis}
As an additional test case, we use reanalysis and altimetry observations of atmospheric predictors to predict altimetry-observed precipitation levels. The dataset consists of hourly snapshots within a $2^\circ\times2^\circ$ box along swaths measured by the Tropical Rainfall Measuring Mission (TRMM) satellites from 2000--2010 \cite{kummerow2000status}. As predictors from ERA5, we use equivalent potential temperatures averaged over the planetary boundary layer (1000--900mb) and free troposphere (850--400mb), as well as column relative humidities and precipitable water integrated over the planetary boundary layer and the free troposphere. Additionally, we use convective available potential energy, surface air temperatures, 500-hPa vertical velocities, and entire-troposphere precipitable water from ERA5. Finally, the subgrid percentage of convective and stratiform areas measured by TRMM radar are used, as they are a leading indicator of precipitation amount \cite{ahmed2015convective}. The first 80\% of observations are used for training (176,224 samples), while the validation and testing set consist of the penultimate and final 10\% of samples, respectively (22,028 samples).

\subsection{Metrics}
\label{subsec:methods.metrics}
We employ several metrics to assess our predictions:
\begin{enumerate}
    \item \textit{Discrimination metrics}. The Mean Absolute Error (MAE) between the true quantiles at probability level $q$ ($\mathbf{y}_q$) and prediction ($\mathbf{\hat{y}}_q$) is computed by 
    \begin{equation}
    \textrm{MAE}(\mathbf{y}_q, \mathbf{\hat{y}}_q) = \frac{1}{n}\sum_{i=1}^n |y_{q,i} - \hat{y}_{q,i}|.
    \end{equation}
    A low quantile prediction MAE implies that forecasts correctly approximate the conditional probability distribution. However, it requires that the ground truth distribution is available.
    \item \textit{Calibration metrics}. Calibration, also sometimes referred to as reliability, measures the probability of forecasted events against the frequency of those events \cite{dawid1982well, gneiting2007probabilistic}. To measure calibration, the Probability Integral Transform (PIT) \cite{dawid1984present} is useful. The premise of the PIT is that the predicted cumulative distribution $\hat{F}_i$ evaluated at the observed value $y_i$ should follow a uniform distribution. Thus, the histogram of $p_i = \hat{F}_i(y_i)$ should be approximately flat, and the deviations from a flat histogram can be measured against the expected level of deviation for a truly uniform distribution. The prediction of different quantiles also lends itself naturally to computing PIT histograms. Consider $m$ uniform quantile predictions $\hat{y}_{q_1, i}, \dots, \hat{y}_{q_m, i}$ for a given sample $i$. This partitions the real line into $B = m+1$ bins $B_k = [\hat{y}_{q_k, i}, \hat{y}_{q_{k+1}, i})$ for $k=0, \dots, m$, using the convention that $\hat{y}_{q_0, i} = -\infty$ and $\hat{y}_{q_{m+1}, i} = \infty$. Given samples $i=1, \dots, n$, let $n_k$ be the total number of observations $y_i$ that fall in $B_k$. The deviation statistic is given by 
    \begin{equation}
        D = \sqrt{\frac{1}{B} \sum_{k=0}^m \left(r_k - \frac{1}{B}\right)^2},
        \label{eq:pit_deviation_statistic}
    \end{equation}
    where $r_k = \frac{n_k}{n}$ gives the frequency of the $k$\textsuperscript{th} bin. This deviation statistic is measured against the expected level of deviation for a true uniform distribution \cite{bourdin2014reliable}
    \begin{equation}
        \mathbb{E}D = \sqrt{\frac{1 - B^{-1}}{nB}}.
        \label{eq:pit_expected_deviation}
    \end{equation}
    Alternatively, the likelihood of the deviation statistic $D$ can be quantified under the null hypothesis that the PIT histogram is sampled from a uniform distribution. Under the limit of large sample size $n$ and bin counts $n_k$, the deviation statistic for samples from a uniform distribution has a chi-squared distribution \cite{wasserman2004all}:
    \begin{equation}
        nB^3D^2 \sim \chi^2(B-1).
    \end{equation}
    Thus, the deviation statistic can be used to test the null hypothesis of uniformity against the alternative that the PIT histogram is nonuniform (and that the predicted distributions are poorly calibrated).
    \item \textit{Proper scores}. We also consider the continuous ranked probability score (CRPS; \citeA{matheson1976scoring}). The CRPS measures the quality of a probabilistic prediction $\hat{F}$ against an observed value $y$. The CRPS is given by 
    \begin{equation}
        \mathrm{CRPS}(\hat{F}, y) = \int_{-\infty}^\infty \left(\hat{F}(x) - \mathbb{1}_{\{x > y\}} \right)^2 \ dx.
        \label{eq:crps_sample}
    \end{equation}
    The CRPS is a strictly proper scoring rule, meaning that on expectation, it is optimal when $\hat{F}$ is the true sampling distribution for the observations $y$ \cite{gneiting2007strictly}. Thus, the CRPS cannot be improved by hedging for alternative outcomes. Predictions are penalized for being overconfident as well as underconfident; as such, the CRPS assesses both the calibration and the sharpness of the predicted distributions simultaneously. Notably, the CRPS is a generalization of the Mean Absolute Error in the case in which predictions are point estimates and $\hat{F}(x)$ is a Heaviside function. The CRPS can be approximated directly from the predicted quantiles by using the quantile formulation of the CRPS from \citeA{laio2007verification} and employing numerical approximations \cite{brocker2012evaluating, taggart2023estimation}. Under equispaced quantile predictions, this is evaluated as
    \begin{equation}
        \mathrm{CRPS}(\mathbf{\hat{y}}_q, y) = \frac{2}{m} \sum_{j=1}^m q (y - y_{q_j})\mathbb{1}_{\{y > y_{q_j}\}} + (1-q) (y_{q_j} - y)\mathbb{1}_{\{y \leq y_{q_j}\}},
    \end{equation}
which is simply double the average pinball loss of the observation under the given quantiles.
\end{enumerate}

\subsection{Optimization and training procedure}
\label{subsec:methods.training}
For an appropriate comparison between the RBLQNN and other baselines, similar architectures and hyperparameter configurations are used for all models for each dataset considered in this study. The hyperparameter configurations for each dataset are given in Supporting Table~\ref{supp:tab:hparam_config}. A hyperparameter sweep over different learning rates, model sizes, and regularizations for the synthetic datasets (Fig.~\ref{fig:training_stability}) indicated that models converged well under similar configurations for the RBLQNN and baselines, justifying the use of identical learning rates and network sizes for comparison of different model types. Model weights are optimized using the Adam optimizer \cite{kingma2014adam}. Weights are saved as checkpoints at the end of each epoch during training, and the checkpoint with the best loss over the validation set is used for analysis on the test set. Early stopping is also employed to limit computational costs from training many thousands of networks.

MVE networks are known to have some stability issues due to gradients typically being more sensitive to the predicted standard deviation than the predicted mean \cite{sluijterman2024optimal, seitzer2022on}. Therefore, one recommendation from \citeA{nix1994estimating} is to employ a ``warm-up period'' where only the mean is optimized while the variance is fixed over the first few epochs. To enforce training stability, during the initial warmup phase a fixed variance $\sigma^2_F$ is prescribed in equation \eqref{eq:gaussian_nll}, and outputted predicted log-variances are penalized for deviations from this fixed variance using an MSE loss:
\begin{equation}
    \mathcal{L}\left(y_i, \hat{\mu}(\boldsymbol{x}_i), \hat{\sigma}^2(\boldsymbol{x}_i)\right) = \frac{1}{2} \left[\left(\frac{y_i - \hat{\mu}_i}{\sigma_F}\right)^2\right] + (\log{\hat{\sigma}_i^2} - \log{\sigma_F^2})^2,
\end{equation}
where the first term is taken from the Gaussian log-likelihood loss \eqref{eq:gaussian_nll} and the second term penalizes predicted log-variances, respectively.

The RBLQNN was found to converge more stably than the MVE networks without any such warm-up period. However, including an analogous warm-up phase for the RBLQNN tended to result in sharper conditional probability estimates. Thus, for the RBLQNN, during the same initial warm-up epochs used for the MVE networks, an MSE loss is applied between the RBLQNN's outputs and the training targets. After the warm-up period is complete, the weights are adjusted using the loss \eqref{eq:net_quantile_loss} to predict the quantiles of the distribution.

A summary of the hyperparameter configurations for each dataset is given in Supporting Table~\ref{supp:tab:hparam_config}. 

\section{Model performance on synthetic data}
\label{sec:synthetic_results}

\subsection{Model error}
\label{subsec:synthetic_results.errors}

\begin{figure}[!ht]
    \centering
    \includegraphics[width=\linewidth]{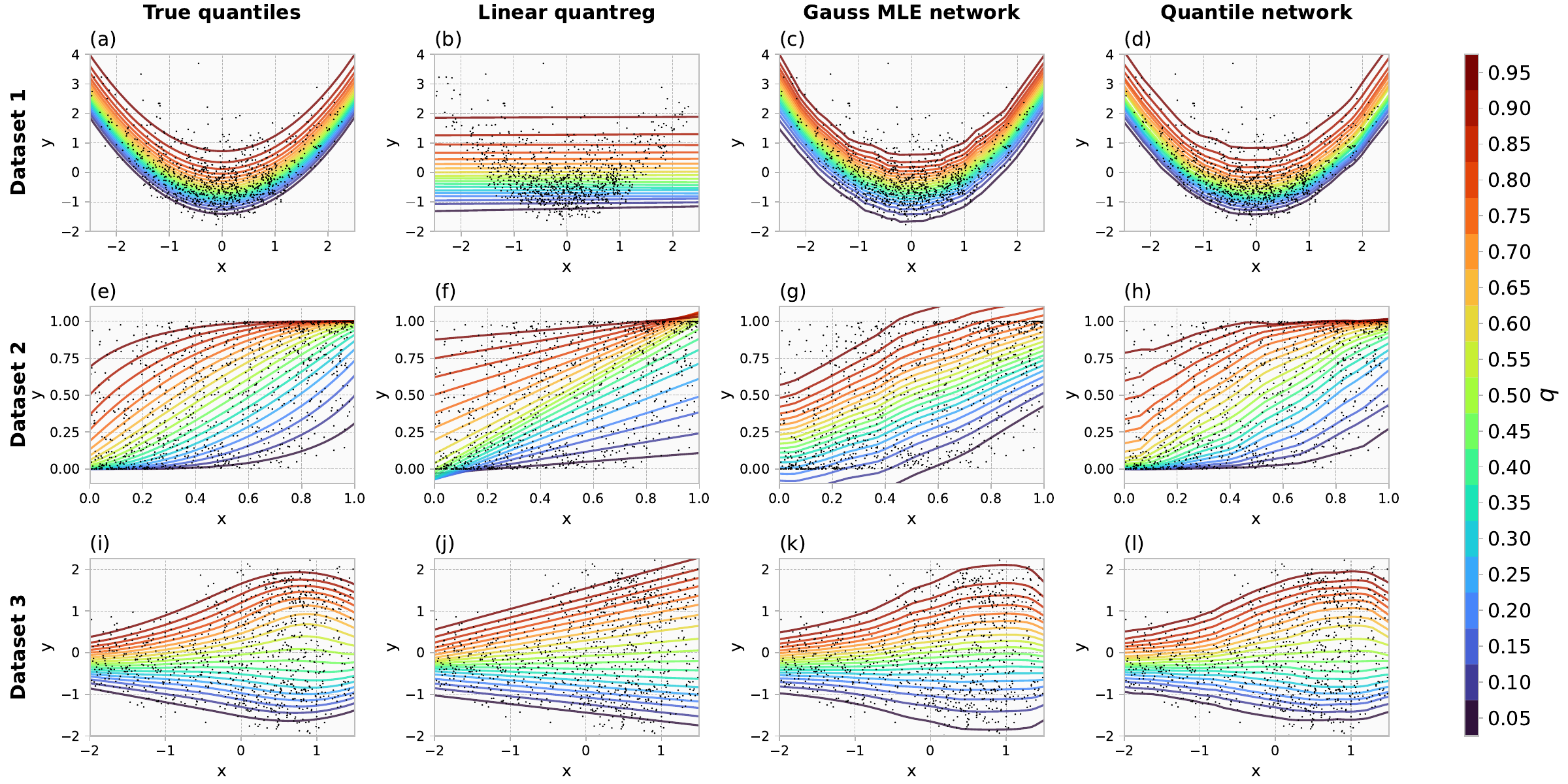}
    \caption{Predicted conditional quantiles on Synthetic Dataset 1 (a--d), Dataset 2 (e--h), and Dataset 3 (i--l). Scatterplot shows the test sample generated from the ground truth distribution, while colored lines indicate predicted quantiles at specified probability levels $q$. (a, e, i) True quantiles. (b, f, j) Predictions made using linear quantile regression. (c, g, k) Predictions made using MVE networks. (d, h, l) Predictions made using the RBLQNN.}
    \label{fig:toy_data_quantile_predictions}
\end{figure}

Figure \ref{fig:toy_data_quantile_predictions} shows the predicted quantiles of the different conditional probability estimation techniques on the three synthetic datasets described in Section~\ref{subsubsec:methods.data.synthetic}. Signed errors $\hat{y}_q - y_q$ in the quantile predictions as a function of $x$ are given in Supporting Figure~\ref{supp:fig:quantile_errors_by_x}. These datasets clearly illustrate the limitations of LQR and the MVE neural networks: LQR fails to estimate highly nonlinear functional dependencies (e.g. Figure~\ref{fig:toy_data_quantile_predictions}b), while MVE networks fail to capture non-Gaussian conditional distributions (Fig.~\ref{fig:toy_data_quantile_predictions}g). However, even when the assumptions of linearity or Gaussianity provide decent but imperfect approximations, the RBLQNN yields better estimates of the conditional distributions. For instance, although the normal distribution and Gumbel distribution share certain features such as unimodality and infinite support (Fig.~\ref{fig:toy_data_quantile_predictions}a), the RBLQNN is able to capture the skewness of the distribution (Fig.~\ref{fig:toy_data_quantile_predictions}d) while the MVE neural network cannot (Fig.~\ref{fig:toy_data_quantile_predictions}c). Similarly, although the linear quantile regression accurately approximates the nearly linear median quantiles of Dataset 2 (Fig.~\ref{fig:toy_data_quantile_predictions}e and \ref{fig:toy_data_quantile_predictions}g), the quantiles at the tails of the distribution are poorly predicted by the linear quantile regression model while they are well-fit for the RBLQNN (Fig.~\ref{fig:toy_data_quantile_predictions}h). 

\begin{figure}[!ht]
    \centering
    \includegraphics[width=\linewidth]{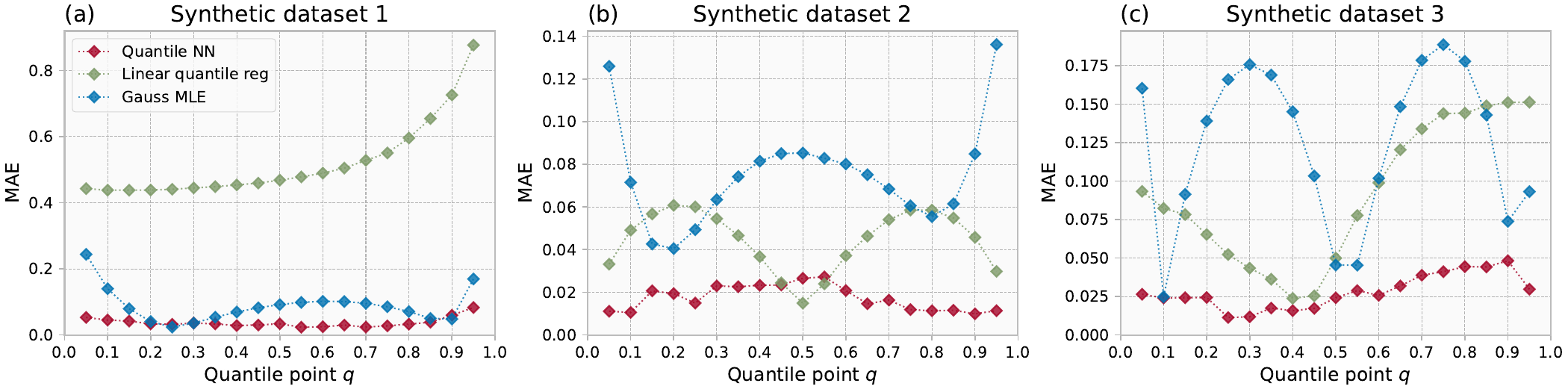}
    \caption[Mean absolute errors of predicted quantiles for the RBLQNN, linear quantile regression, and MVE neural network for synthetic datasets]{Mean absolute errors of predicted quantiles for the RBLQNN (red), linear quantile regression (light green), and MVE neural network (blue) for (a) Synthetic Dataset 1 and (b) Synthetic Dataset 2.}
    \label{fig:toy_data_quantile_mae}
\end{figure}

Figure \ref{fig:toy_data_quantile_mae} shows the average errors of each of the individual quantile predictions averaged over the test set and further demonstrates how the assumptions made by each baseline result in poorly predicted distributions. Dataset 1 shows that inadequately capturing nonlinear dependencies with linear quantile regression can result in systematically large errors for all quantiles in the predicted conditional distribution (Fig. \ref{fig:toy_data_quantile_mae}a). The MVE network resolves the issue of nonlinearity, but the assumption of Gaussianity can lead to poor estimation of certain quantiles. For instance, while the MVE network provides reasonable estimates for the bulk of the distribution in Dataset 1, the tails of the distribution tend to be underestimated (Fig.~\ref{fig:toy_data_quantile_predictions}c and Fig.~\ref{fig:toy_data_quantile_mae}a). Large errors in the tails of the distribution also occur in Dataset 2 and 3, as the Gaussian approximation is unable to capture the compact support of the distribution for Dataset 2 nor the bimodality of Dataset 3. The bimodality of Dataset 3 also results in the MVE network producing poor quantile predictions near each lobe of the bimodal conditional distribution (Fig.~\ref{fig:toy_data_quantile_predictions}k and \ref{fig:toy_data_quantile_mae}c). In contrast to these baselines, the RBLQNN tends to produce quantile predictions that are good across all quantiles of the distribution.

\subsection{Comparison against other quantile regression neural network approaches}
\label{subsec:synthetic_results.qnn_comparison}

\begin{figure}[!ht]
    \centering
    \includegraphics[width=\linewidth]{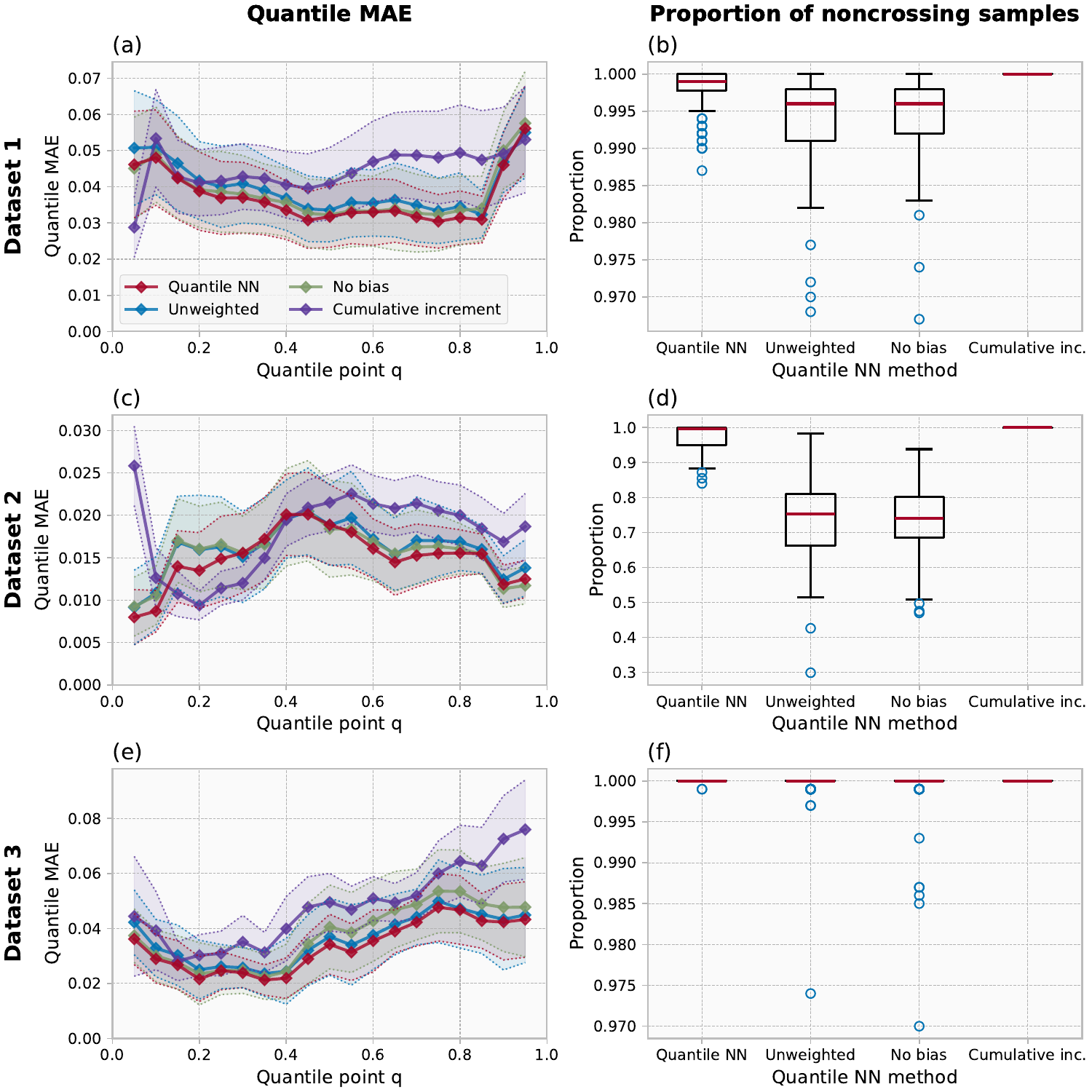}
    \caption{Performance of different quantile regression neural network techniques on Synthetic Dataset 1 (a, b), Dataset 2 (c, d), and Dataset 3 (e, f). (a, c, e): Mean absolute errors for quantile neural network predictions, averaged over 100 ensembles created by initializing network weights with different random seeds. Red: RBLQNN (Eq. \ref{eq:net_quantile_loss}). Blue: Unweighted quantile regression neural network of \citeA{xu2017composite}. Light green: No-bias quantile regression neural network framework implementing the inverse-expectation weighting but without the ReLU bias loss (Eq.~\ref{eq:relu_bias_loss}). Purple: cumulative increment network of \citeA{padilla2022quantile}. (b, d, f) Boxplots showing the fraction of nondegenerate probability distributions predicted by each quantile neural network approach. Conditional probability distributions are considered degenerate for a given sample if the predicted quantiles are not monotonic.}
    \label{fig:toy_data_qnn_comparison}
\end{figure}

Figure \ref{fig:toy_data_qnn_comparison} compares the performance of the RBLQNN to the alternative quantile regression neural network approaches given in Section~\ref{subsubsec:methods.conditional_probability.baselines}. Since the network performance is sensitive to the weight initialization, for each quantile neural network method, an ensemble of 100 networks with weights initialized using different random seeds is used to robustly evaluate performance. 

For the three synthetic datasets, the RBLQNN tends to produce better quantile predictions than the baseline quantile neural network techniques (Fig.~\ref{fig:toy_data_qnn_comparison}a, \ref{fig:toy_data_qnn_comparison}c, and \ref{fig:toy_data_qnn_comparison}e). The RBLQNN error is similar to, though on average slightly lower than, the quantile neural networks which directly predict quantiles (the ``unweighted" and ``no-bias" networks). This seems to be primarily due to the inclusion of the ReLU bias loss (Eq.~\ref{eq:relu_bias_loss}), as the comparisons between the two baselines with different quantile weighting schemes do not result in substantially different performance. Including the ReLU bias loss to encourage predicted distributions to be valid may impose stability constraints which facilitate network convergence during training.

Figure~\ref{fig:toy_data_qnn_comparison}b, \ref{fig:toy_data_qnn_comparison}d, and \ref{fig:toy_data_qnn_comparison}f show the proportion of test samples which result in monotonic quantile predictions (i.e., nondegenerate conditional probabilities). The RBLQNN results in significant reductions in the number of predicted conditional distributions that are degenerate when compared to the unweighted and no-bias networks. For instance, applying the ReLU bias loss in Dataset 2 reduces the average number of samples with quantile crossings from more than 25\% to less than 3\% (Supporting Table~\ref{supp:tab:avg_quantile_crossings}). For Dataset 3, all but two ensemble members completely eliminate quantile crossings with the RBLQNN, whereas roughly 20\% of ensemble members have quantile crossings for the unweighted and no-bias quantile networks (Supporting Table~\ref{supp:tab:prop_perfect_distributions}).

While the RBLQNN reduces the number of quantile crossings compared to other quantile neural networks which directly predict quantiles, it does not completely eliminate them. In contrast, the cumulative increment network's design explicitly prohibits all quantile crossings. However, Figures~\ref{fig:toy_data_qnn_comparison}a, \ref{fig:toy_data_qnn_comparison}c, and \ref{fig:toy_data_qnn_comparison}e all show that the predicted distributions made using the cumulative increment network tend to be worse than the quantile neural networks which directly output quantiles. The relatively poorer performance of the cumulative increment network is likely due to instabilities resulting from the sum in \eqref{eq:cuminc_output} used for predicting higher order quantiles.

\subsection{Training stability}
\label{subsec:synthetic_results.training_stability}

A central challenge in implementing neural networks is that model performance can be sensitive to various hyperparameters \cite{goodfellow2016deep}. Weights and biases in a neural network are normally optimized using stochastic gradient-based optimization techniques, which can evolve unpredictably under the highly nonconvex, high-dimensional loss landscapes set by the training data, network architecture, regularization, and loss functional. The convergence of a neural network thus depends on all of the factors that determine the loss landscape as well as the specifications that prescribe the optimization procedure. As a result, hyperparameter selection often requires extensive tuning or automated sweeps over various configurations to obtain a well-fitted model. It is therefore advantageous to train using a loss function that converges robustly across a range of hyperparameters and datasets. We investigate the sensitivity of the RBLQNN to MVE networks across the three synthetic datasets.

Network training is often particularly sensitive to hyperparameters such as the learning rate, network architecture, and regularization \cite{smith2018disciplined,godbole2023tuning}. Therefore, we perform a grid search over different hyperparameters, including learning rates, regularizations, and network sizes, with values provided in Supporting Table~\ref{supp:tab:training_stability_hparams}. We use eight different learning rates, three regularization levels, 2 layer sizes and 3 different numbers of neurons per layer, to sample a total of 144 hyperparameter combinations. All other hyperparameters are set using the configuration given in Supporting Table~\ref{supp:tab:hparam_config}.

\begin{figure}[!ht]
    \centering
    \includegraphics[width=\linewidth]{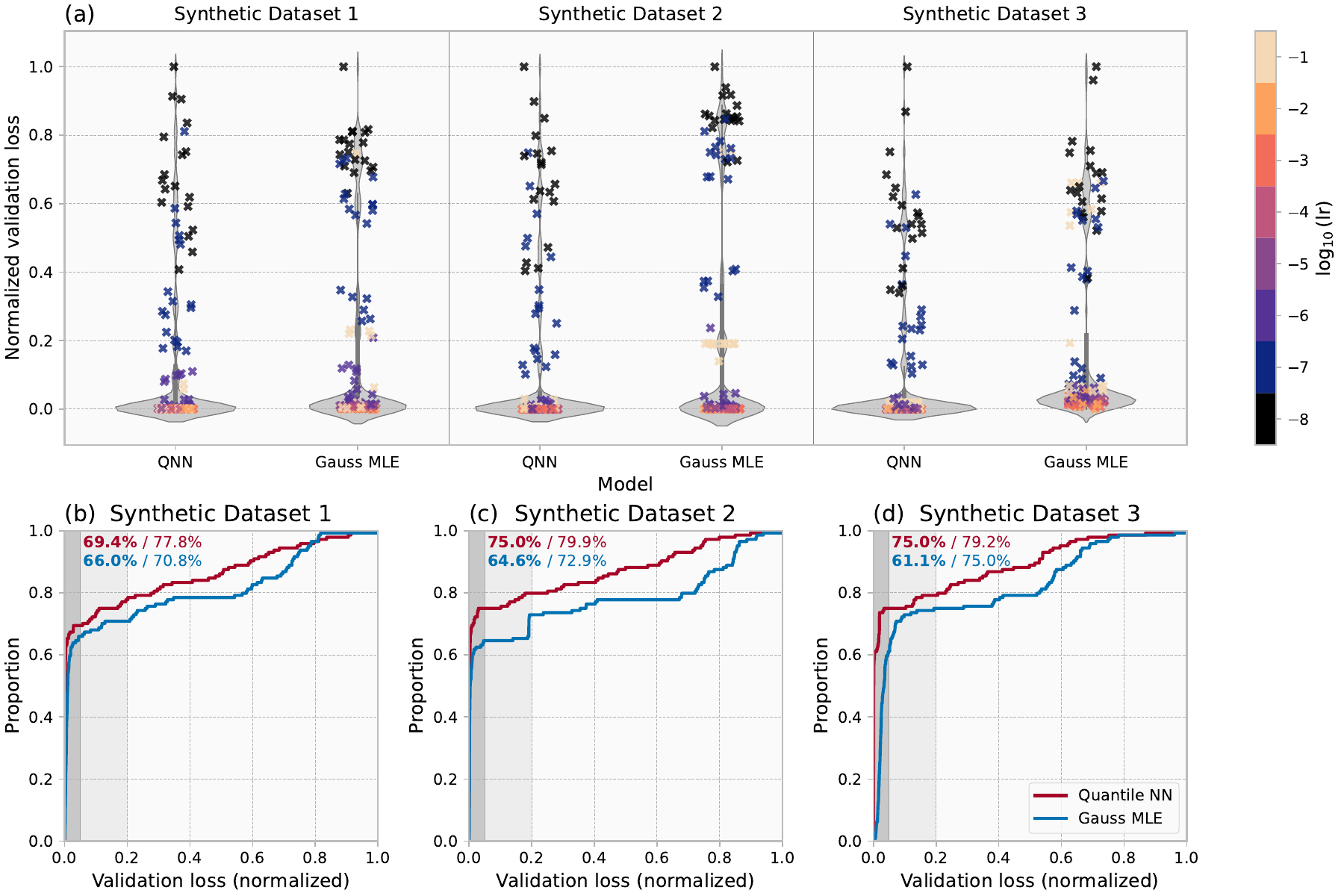}
    \caption{Training stability of the RBLQNN and MVE network evaluated over 144 hyperparameter combinations for the three synthetic datasets. (a) Strip plots of the lowest validation loss attained during training for the RBLQNN and MVE for the three different datasets, with gray shading showing a kernel density estimate. Losses are normalized to the $[0, 1]$ range. Colors indicate the learning rate. (b, c, d) Empirical CDFs of the normalized validation losses for each of the three different synthetic datasets for the RBLQNN (red) and MVE network (blue). Boldface percentages in the top left corner indicate proportion of neural networks which converge within 5\% of the best loss from all hyperparameter configurations (indicated by dark gray shading). Plain typeface percentages indicate proportions of neural networks within 20\% of the best loss (light gray shading).}
    \label{fig:training_stability}
\end{figure}

Figure~\ref{fig:training_stability}a shows the best validation loss attained during training by the RBLQNN and MVE network for each of the 144 hyperparameter combinations. To permit a comparison between the different loss functions for the different networks, validation losses are normalized to the $[0, 1]$ range so that the hyperparameter configuration with the lowest loss over the 144 hyperparameter combinations has a normalized value of 0 and the highest loss has a normalized value of 1. Kernel density estimates indicate that losses cluster near the minimum values, indicating convergence for a broad range of hyperparameters for both the RBLQNN and MVE networks. However, for each of the three datasets, more values cluster near the minimum loss for the RBLQNN than for the MVE network, suggesting that training tends to be more stable for the RBLQNN than for the MVE network. Notably, the RBLQNN appears to attain strong performance for a broader range of learning rates, a crucial hyperparameter for neural network training. For low learning rates (e.g. $10^{-8}$), normalized validation losses are high for each network; however, increasing learning rates results in better performance gains for the RBLQNN than for the MVE network (e.g., a normalized validation loss that is 18.2\% lower on average for the RBLQNN than MVE network when using learning rates of $10^{-7}$). For high learning rates ($10^{-1}$), losses begin to increase significantly for the MVE network but less so for the RBLQNN.

Figure~\ref{fig:training_stability}b, \ref{fig:training_stability}c, and \ref{fig:training_stability}d show the empirical cumulative distribution function of the losses given in the strip plots in Figure~\ref{fig:training_stability}a for each of the three synthetic datasets. The CDF of the RBLQNN is above the CDF of the MVE network at most loss levels, indicating that a greater proportion of RBLQNN converge within a given margin of the lowest loss than the MVE networks. For instance, for Synthetic Dataset 1, only 66\% (71\%) of MVE neural networks converge within 5\% (20\%) of the minimum loss, whereas 69\% (78\%) of the RBLQNNs converge within this margin of the minimum loss.

\subsection{Sample-based metrics}
\label{subsec:synthetic_results.crps_calibration_sharpness}

The low MAE of the predicted quantiles made on the synthetic datasets indicates that the RBLQNN can successfully approximate conditional distributions. However, for most datasets, the ground truth distribution is unknown, making it impossible to assess predicted conditional probabilities directly against the true distribution. In this section, we explore sample metrics that assess the predicted distributions against the data.

\begin{figure}[p]
    \centering
    \includegraphics[width=0.8\linewidth]{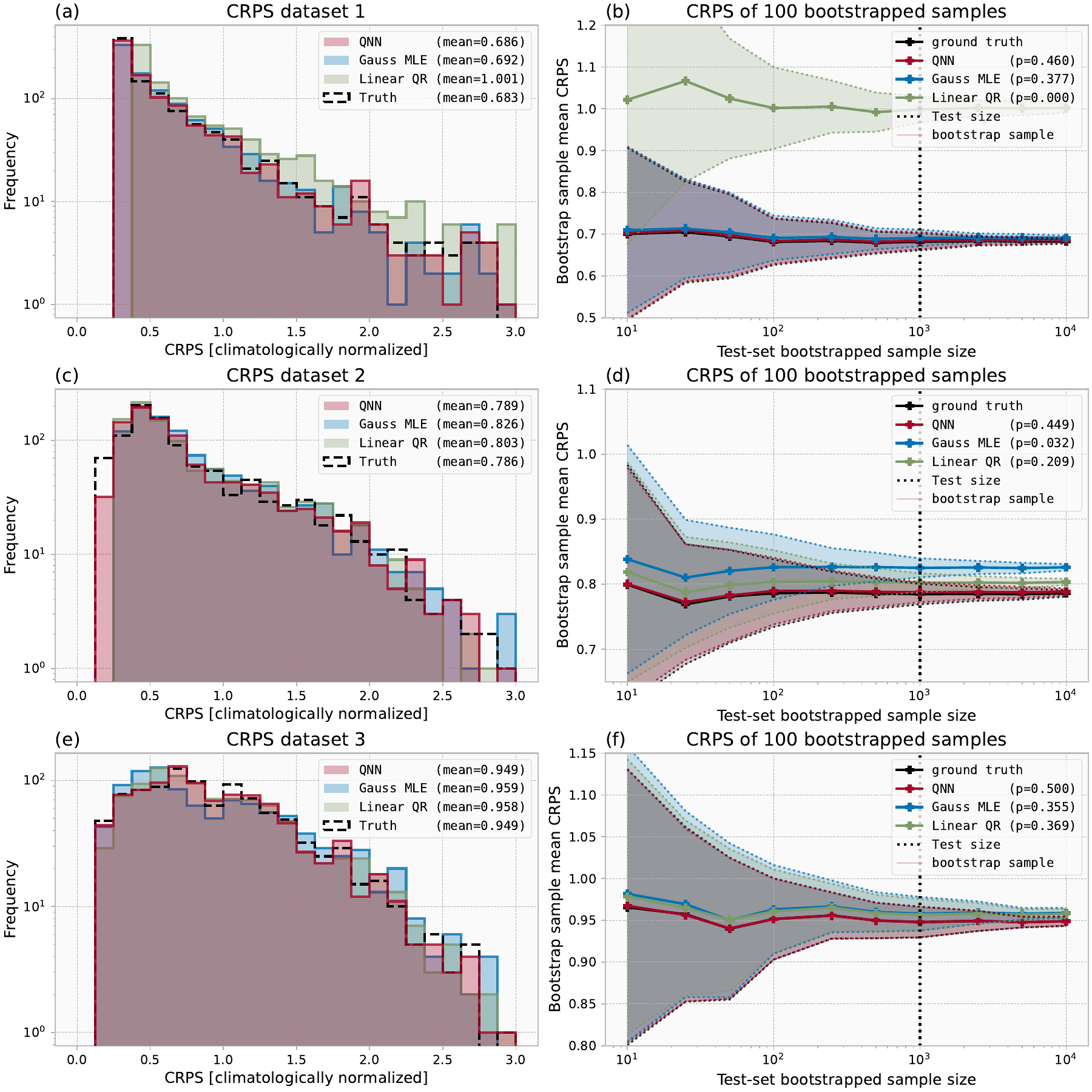}
    \caption{Continuous ranked probability score evaluated over the test samples of Synthetic Dataset 1 (a, b), Dataset 2, (c, d) and Dataset 3 (e, f). (a, c, e) Histograms of the continuous ranked probability score yielded by the predicted distributions of the RBLQNN (red), MVE network (blue), and linear quantile regression (light green) evaluated over the 1,000 test samples. The black dashed line shows the histogram of the CRPS attained by evaluating the ground truth distribution against the observed samples. The continuous ranked probability scores have been normalized relative to the climatological CRPS. Quantities in the legend indicates the sample-average CRPS for each method. (b, d, f) Bootstrap-sampled CRPS sample averages. For a variety of quasi-logspaced sample sizes $N_b$ ranging from $10$ to $10^4$ (multiples of 1, 2.5, 5 times powers of 10), a bootstrapped sample of size $N_b$ is sampled with replacement from the test set, and the CRPS sample average is taken for each method of conditional probability estimation. This is repeated 100 times to create a bootstrap ensemble. The thick notched lines show the ensemble-mean sample-average CRPS for each sample size $N_b$, and the shaded region shows the ensemble spread ($\pm$ one standard deviation). The black dotted line indicates the sample size of the test set ($N=1,000$). The fractions in the legend evaluate the probability that the sample-averaged CRPS for each method will be lower than the CRPS yielded by the ground truth distribution through pairwise comparisons of bootstrapped samples with $N_b=N$. For instance, a fraction of $p=0.03$ for the MVE network (panel d) indicates that only 3\% of bootstrapped CRPS sample averages for the MVE network are lower than the CRPS sample averages computed using the ground truth distribution. This percentage is computed over the 10,000 pairwise comparisons between the 100 ensemble members for the MVE network CRPS sample averages and 100 ensemble members for the ground truth distribution CRPS sample averages.}
    \label{fig:toy_data_crps}
\end{figure}

Figures~\ref{fig:toy_data_crps}a, \ref{fig:toy_data_crps}c, and \ref{fig:toy_data_crps}e show histograms of the CRPS yielded by the predicted distributions of the different conditional probability estimation techniques over the test set. Averaged over all test samples, the RBLQNN has the lowest average CRPS of the three conditional probability estimation techniques for all synthetic datasets. The sample average CRPS for the RBLQNN nearly matches the CRPS obtained by using the ground truth conditional probability distribution for all three datasets.

Despite the RBLQNN attaining a lower sample-averaged CRPS for all three datasets, the histograms of CRPS significantly overlap for each of the different methods. Since the CRPS is a sample-based metric which is optimal only on expectation, the statistical significance of the low sample-averaged CRPS must be assessed. To assess the statistical significance of the sample-averaged CRPS over a range of different sample sizes, we employ bootstrapping with 100 ensemble members for a variety of bootstrap sample sizes $N_b$ from 10 to $10,000$. For each ensemble member, $N_b$ samples are drawn with replacement from the test set, and the sample-averaged CRPS is computed over that sample for the various conditional probability estimation techniques. The CRPS sample averages for different ensemble members are compared pairwise with the true distribution CRPS sample averages to evaluate whether the predictions made by a given method are statistically distinguishable from the true distribution for a given sample size $N_b$. For instance, if among the 10,000 pairwise comparisons between the 100 ensemble members of the MVE network and 100 members of the true distribution CRPS sample averages fewer than 5\% of the comparisons yield lower CRPS sample means for the MVE network, then it can be concluded that the probabilistic predictions made by the MVE network are inadequate for that sample size.

Figures~\ref{fig:toy_data_crps}b, \ref{fig:toy_data_crps}d, and \ref{fig:toy_data_crps}f show the ensemble spread of the CRPS sample averages for each of the different methods for a variety of bootstrap sample sizes $N_b$. As the bootstrap sample size increases, the ensemble spread of the CRPS sample averages decreases for each of the different methods, revealing the minimum sample sizes needed to reject the probabilistic models. For instance, for Dataset 1 only a few hundred samples are needed to establish that the linear quantile regression poorly predicts probability distributions, whereas $1,000$ samples are needed to reject the probabilistic predictions made by the MVE network for Dataset 2. In contrast, the probability distributions predicted by the RBLQNN cannot be rejected up to samples of size $10,000$. The evaluations of the CRPS in Figure~\ref{fig:toy_data_crps} indicate that even in cases in which the predicted probability distributions are superior for a given method, substantial sample sizes may be needed to discern performance using sample-based metrics such as the CRPS. Care should therefore be taken when interpreting differences between sample-mean CRPS for different methods, especially for datasets with small sample sizes.

\section{Model performance on observational datasets}
\label{sec:observational_results}

We next evaluate the performance of the RBLQNN against the MVE network and linear quantile regression for the GSOD daily maximum temperature datasets as well as the TRMM precipitation dataset. For the GSOD datasets, we evaluate models trained separately at 1,501 NOAA weather stations, with 1,107--1,827 test samples spanning 2016--2020. The TRMM dataset is a single dataset with $22,028$ test samples. As the ground truth probability distribution is unknown for these datasets, performance is evaluated using sample-based metrics such as the CRPS and PIT histogram deviation statistic.

\subsection{GSOD daily maximum temperatures}
\label{subsec:observational_results.gsod}

\begin{figure}[p]
    \centering
    \includegraphics[width=0.85\textwidth]{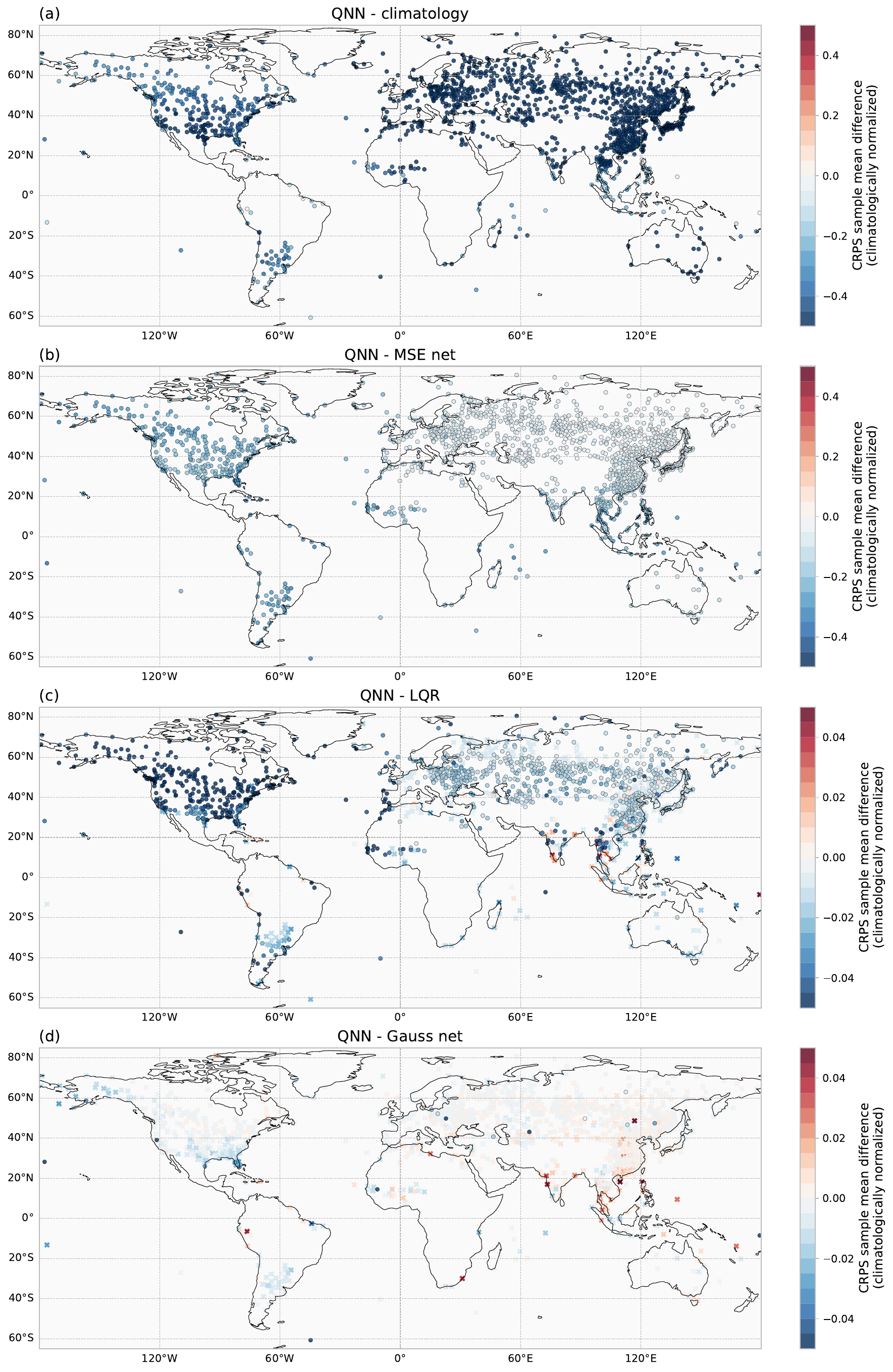}
    \caption{Comparison of sample-average CRPS between RBLQNN predictions and baselines. Maps of the difference in sample-mean CRPS between RBLQNN and (a) climatology, (b) MSE network, (c) linear quantile regression, and (d) mean-variance estimation network. Values have been normalized with respect to the climatological CRPS sample mean at each location. Circles outlined in black indicate statistically significant differences, determined using the pairwise comparisons method of Section~\ref{subsec:synthetic_results.crps_calibration_sharpness}. Crosses indicate differences which are not statistically significant. Note the order of magnitude difference in the colorbar extent for panels (a, b) vs (c, d).}
    \label{fig:gsod_crps_difference}
\end{figure}

Figure~\ref{fig:gsod_crps_difference} compares the sample-averaged CRPS of the RBLQNN against the various baselines. Histograms of the sample-averaged CRPS over all locations for each of the different methods are shown in Supporting Figure~\ref{supp:fig:gsod_crps_histogram}. To maintain consistency across locations with different climatological variability, CRPS values are normalized by the CRPS obtained from applying the climatological quantiles of daily maximum temperatures at each location, so that a normalized sample-mean CRPS of 1 indicates probabilistic predictions no better than climatology. For all three conditional probability estimation techniques, time-mean CRPS is lower than the climatological CRPS at 1,493 out of 1,501 stations (99.4\%). Moreover, the probabilistic models result in better CRPS than the MSE network at 1,492 of 1,401 locations (99.2\%). Thus, all three conditional probability estimation techniques tend to provide improved information about the conditional distribution, which is not permitted by either climatological or deterministic baselines. 

The differences in CRPS between the RBLQNN and other conditional probabilistic baselines (Fig.~\ref{fig:gsod_crps_difference}c and \ref{fig:gsod_crps_difference}d) are less pronounced than the difference in CRPS for the RBLQNN and climatological or MSE network predictions (Fig.~\ref{fig:gsod_crps_difference}a and \ref{fig:gsod_crps_difference}b). Sample-mean CRPS for the RBLQNN is lower than the CRPS for the linear quantile regression at 1,477 of 1,501 locations (94.8\%). The locations where the linear quantile regression outperforms the RBLQNN could indicate overfitting to out-of-sample distributions, or random variation due to insufficient statistical significance arising from an insufficient sample size. Using pairwise comparisons of bootstrapped ensemble members to assess significance as in Section~\ref{subsec:synthetic_results.crps_calibration_sharpness} shows that the sample-mean CRPS is statistically significantly lower for the RBLQNN than linear quantile regression at 950 of 1,501 stations (63.2\%), indicating that more samples may be needed to attain statistically significant lower CRPS with the RBLQNN for many locations.

Sample-mean CRPS of the RBLQNN is lower than the MVE at 836 of 1,501 locations (55.7\%). While a slight majority of locations show lower CRPS with the RBLQNN than with the MVE network, only 19 of 1,501 stations (1.3\%) show a statistically significant difference. This can indicate that the conditional probability distributions of TMAX are relatively well described by Gaussian distributions, or that more samples are needed to distinguish the skill of the RBLQNN predictions from that of the MVE network.

\begin{figure}[!ht]
    \centering
    \includegraphics[width=0.9\linewidth]{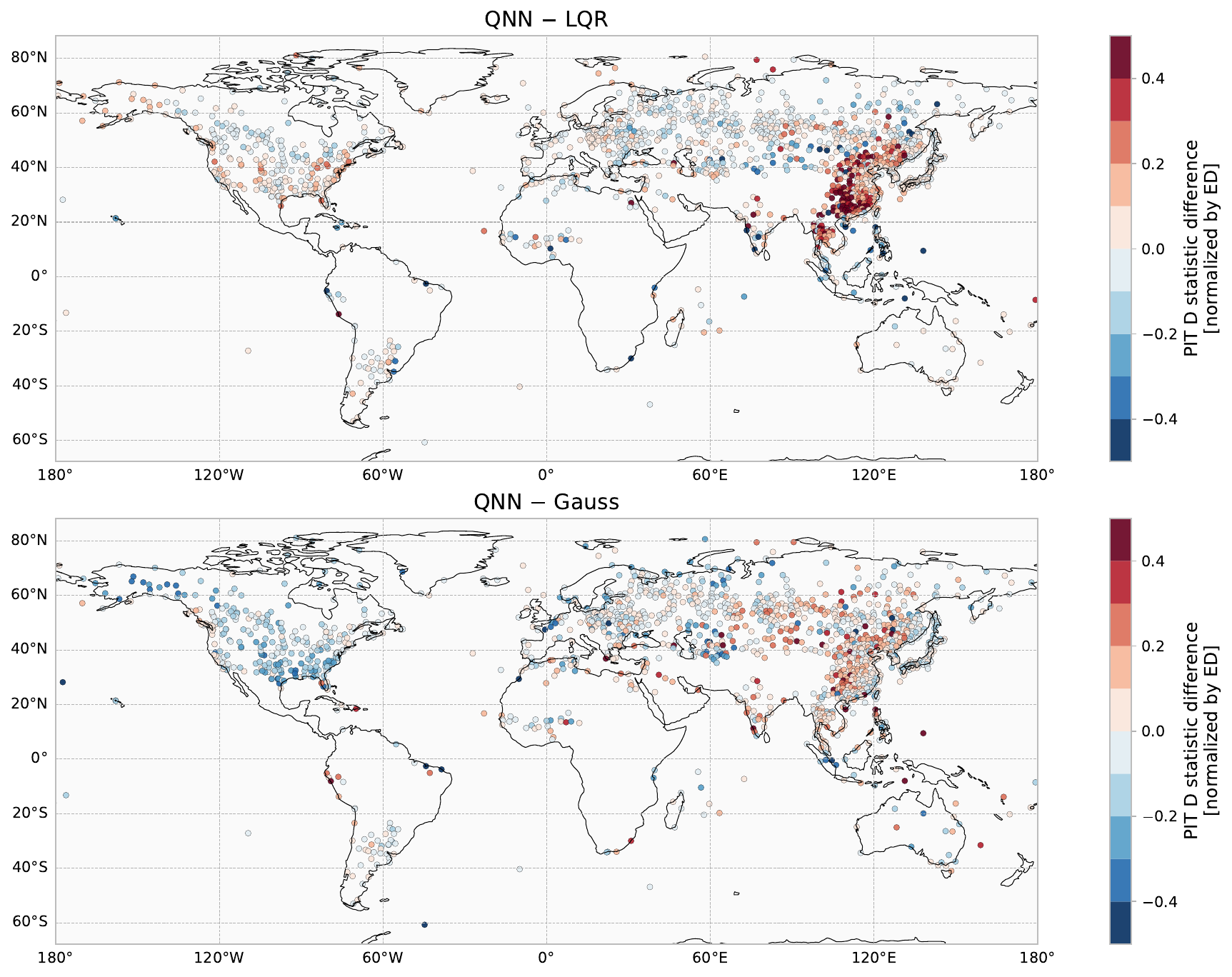}
    \caption{Differences in calibration statistics for RBLQNN predictions and those of (a) linear quantile regression and (b) MVE networks. Calibration is measured using the deviation statistic of the PIT histogram described in Section~\ref{subsec:methods.metrics}. Deviation statistics are normalized by the expected level of deviation for each location.}
    \label{fig:gsod_pit_differences}
\end{figure}

Despite the limitations of the small test sizes for individual stations, a few cohesive geographical regions where the RBLQNN has lower sample-average CRPS than the MVE network point to systematic causes for better predictions from the RBLQNN. The greatest decrease in CRPS from using the RBLQNN instead of the MVE network occurs primarily in the southeastern United States and Alaska. PIT deviation statistics (Fig.~\ref{fig:gsod_pit_differences}) indicate that predictions are relatively well-calibrated for the RBLQNN in these regions. For instance, in North and South America, 278 of 350 stations (79\%) have better calibration statistics for the RBLQNN than the MVE network. Furthermore, while 55 of 350 locations (15\%) have PIT histograms fully consistent with the null hypothesis of uniformity for the RBLQNN, only 11 (3\%) of the stations have PIT histograms consistent with uniformity using the MVE network. The relatively well-calibrated probabilistic predictions of the RBLQNN relative to the MVE network predictions in the southeastern United States suggest that the RBLQNN is successfully estimating inherently non-Gaussian conditional distributions of TMAX. This region is consistent with the regions of high negentropy highlighting non-Gaussian marginal distributions in Figure~\ref{fig:gsod_skewness_kurtosis}c.

Regions where the RBLQNN yields higher CRPS than the MVE network include Southeast Asia (particularly the Malay Peninsula and southeastern China), Siberia, and the Indian subcontinent. In these regions, the MVE network's comparatively stronger performance indicates that the Gaussian approximation adequately represents the underlying conditional distributions. The validity of the Gaussian approximation may be related to the variability in TMAX attributable to the functional dependence on the regressors. Supporting Figure~\ref{supp:fig:gsod_crps_rsq} shows the distribution of differences in CRPS between the RBLQNN and MVE network as a function of the MSE network $R^2$. Since the MVE network and RBLQNN share the same architecture as the MSE network, the $R^2$ yields an estimate about the proportion of variability which is explained by the nonlinear functional dependence of TMAX on the model inputs as opposed to the conditional distribution itself. Supporting Figure~\ref{supp:fig:gsod_crps_rsq}c suggests that the Gaussian approximation is the most valid when a high proportion of the variance is explained by the deterministic functional ($R^2 \approx 1$) or when a negligible proportion of the variance is explained by the deterministic component ($R^2 \approx 0$). In regions such as the Malay peninsula, $R^2$ of the MSE network is low, possibly due to the influence of the monsoon system and effects of moisture variations on the temperature profile as well as complex orography. In such regions, the inputs to the RBLQNN may be rather uninformative, posing challenges for optimizing individual quantiles in the RBLQNN. Conversely, in regions such as southeastern China where $R^2$ is high, much of the variability in TMAX is explained by the deterministic component, and the remaining variability may resemble Gaussian noise. The RBLQNN tends to outperform that of the MVE networks in regions of intermediate $R^2$, in which the inputs are informative yet much of the variance is not fully explained by the deterministic functional. For instance, at stations where the MSE network $R^2$ is between 0.3 and 0.7, 73.1\% of stations have lower time-averaged CRPS with the RBLQNN than with the MVE network.

\begin{figure}[!ht]
    \centering
    \includegraphics[width=\linewidth]{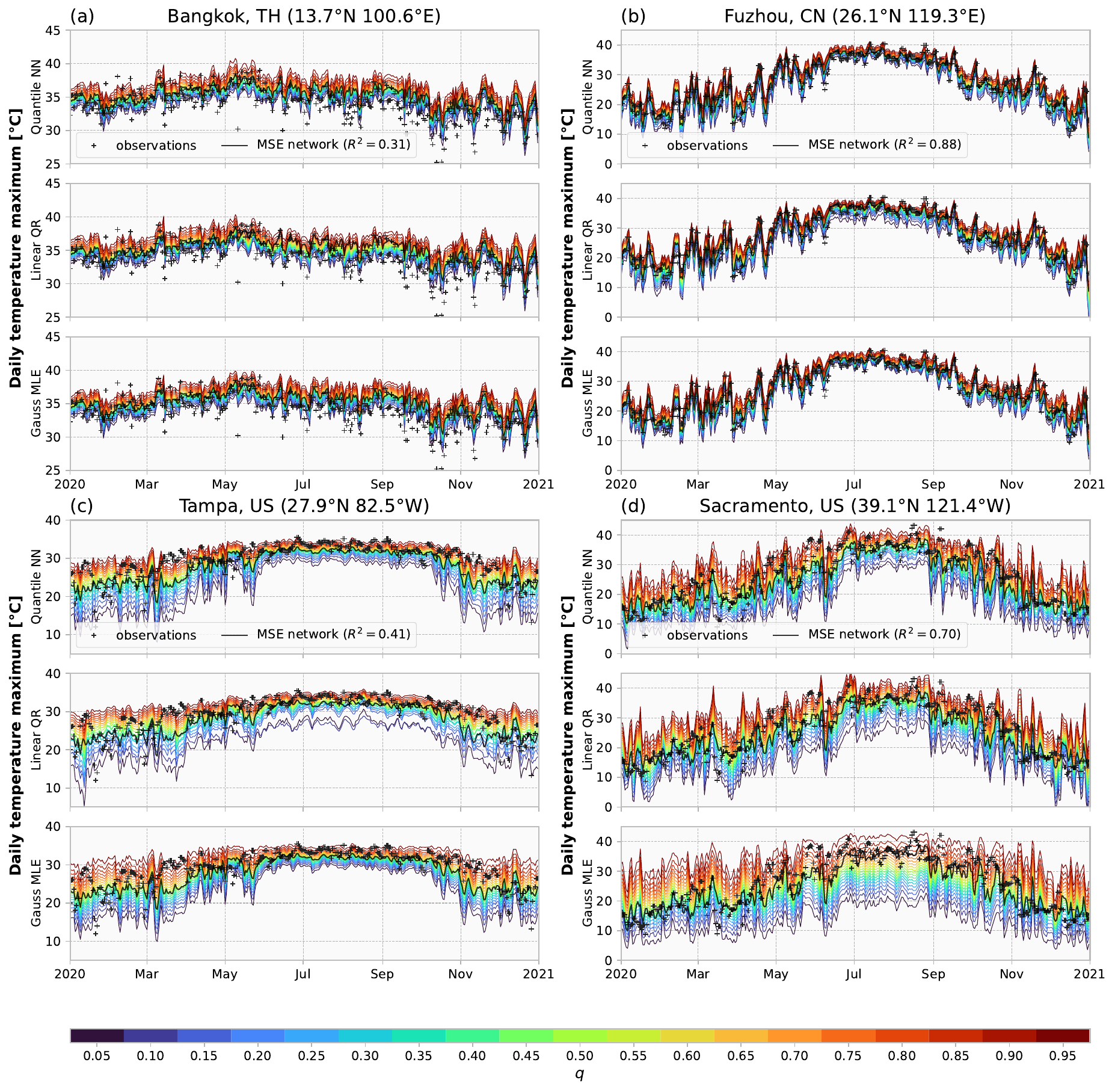}
    \caption{Timeseries of predicted conditional probability distributions for GSOD TMAX for year 2020 in (a) Bangkok, (b) Fuzhou, (c) Tampa, and (d) Sacramento. Colored lines indicate the predicted quantiles, with different panels for RBLQNN (top), linear quantile regression (middle) and mean-variance estimation networks (bottom). Black crosses indicate observed TMAX. The black line indicates predictions made by the MSE network, with $R^2$ given in the legend.}
    \label{fig:gsod_timeseries}
\end{figure}

The timeseries of predicted conditional probability distributions shown in Figure~\ref{fig:gsod_timeseries} help illustrate the situations in which the RBLQNN performs well. Figures~\ref{fig:gsod_timeseries}a and \ref{fig:gsod_timeseries}b show two stations where the RBLQNN has a greater average CRPS than the MVE network. At Bangkok (Fig.~\ref{fig:gsod_timeseries}a), temporal changes in the distribution are small over the course of the year, with the observed sea level pressure and geopotential heights providing minimal information about TMAX. On the other hand, at Fuzhou, (Fig.~\ref{fig:gsod_timeseries}a), nearly all of the variance in temperature is explained by the deterministic component. In both of these cases, the principle of maximum entropy validates the Gaussian approximation, as there is a lack of constraining information about the conditional distributions of TMAX.

Figures~\ref{fig:gsod_timeseries}c and \ref{fig:gsod_timeseries}d show two examples where the RBLQNN does outperform the MVE network, and the differences are statistically significant. The predicted probability distributions respond to seasonality, both through the change in mean predicted temperature throughout the year as well as changes in variability between the winter and summer months. However, the distributions predicted by the quantile neural network are negatively skewed, allowing the RBLQNN to permit cold extremes during the winter while maintaining sharp predicted distributions.

\subsection{TRMM precipitation}
\label{subsec:observational_results.trmm}

While the conditional probability estimates of the RBLQNNs trained to predict temperatures in Section~\ref{subsec:observational_results.gsod} mostly outperform those of linear quantile regression, differences between the RBLQNN and Gaussian maximum likelihood networks are less pronounced. This may be because the probability estimates of TMAX conditioned on sea level pressure and geopotential heights are sufficiently approximated by normal distributions at most stations, or because the test size is insufficient to assess the probabilistic predictions. In this section, we focus on a single dataset which contains significantly more samples (20,028 test samples as opposed to at most 1,827), and for which the Gaussian approximation is clearly less valid.

\begin{figure}[!ht]
    \centering
    \includegraphics[width=\linewidth]{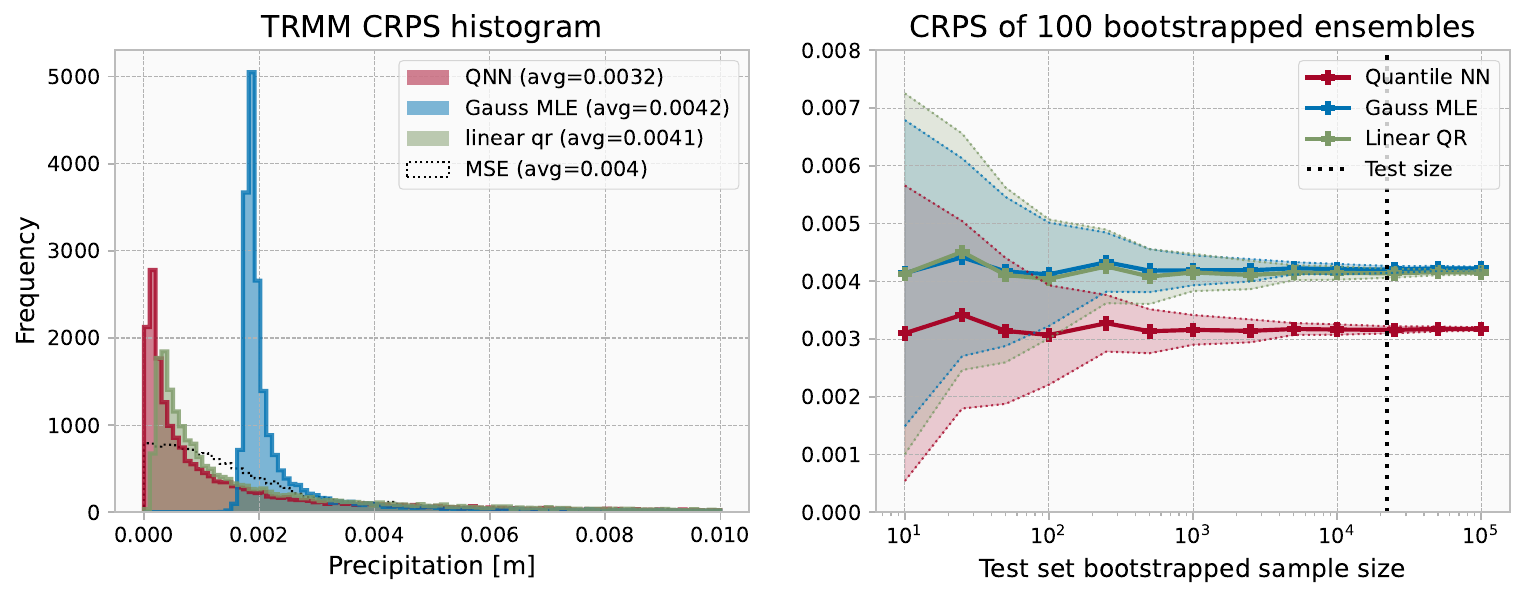}
    \caption{CRPS of different conditional probability estimation techniques evaluated on the TRMM precipitation dataset. (a) Histograms of the CRPS evaluated over all test samples for the RBLQNN (red), Gaussian maximum likelihood network (blue), linear quantile regression (light green) and MSE network (black dotted lines). Values in the legend indicate sample-averaged values. (b) As in Figure~\ref{fig:toy_data_crps}b, \ref{fig:toy_data_crps}d, and \ref{fig:toy_data_crps}f but for the TRMM dataset.}
    \label{fig:trmm_crps}
\end{figure}

The CRPS is shown for the different conditional probability estimation techniques for the precipitation dataset in Figure~\ref{fig:trmm_crps}a. Here, it is clear that the CRPS is substantially better for the RBLQNN than all of the other baselines. Probabilistic predictions are the worst for the Gaussian neural network, and both the Gaussian neural network and linear quantile regression produce probabilistic estimates with even worse scores than an MSE-trained deterministic network. Computing the CRPS using bootstrapped samples (Figure~\ref{fig:trmm_crps}b) shows that for bootstrapped sample sizes of 500 samples or more, the RBLQNN consistently results in better sample-averaged CRPS than the other baselines. Thus, the RBLQNN produces significantly better probabilistic predictions on the precipitation dataset than the Gaussian maximum likelihood network or the linear quantile regression method. This indicates that both nonlinear functional dependence and non-Gaussianity are essential properties of the conditional distributions of precipitation.

\section{Discussion}
\label{sec:discussion}

The chaotic and nonlinear dynamics of the Earth system implies that geophysical variables are prone to fluctuations and uncertainties, posing challenges to predicting geophysical variability with complete certainty. Thus, managing weather and climate risk requires quantifying and constraining estimates for geophysical variability, which depends on fluctuating observable conditions. In this paper, we propose an approach for characterizing the uncertainties of geophysical quantities, such as daily maximum temperatures and precipitation amounts, based on other measurable conditions using quantile regression neural networks. To address some typical issues with quantile neural networks, our implementation---termed the ``ReLU bias loss quantile neural network'' (RBLQNN)---employs two novel, explicit modifications to the loss function to predict quantiles with equal weight and to mitigate the risk of predicting degenerate probability distributions. Using a suite of different datasets---synthetic distributions, in-situ daily temperature maxima observations from weather stations, and altimetry-observed precipitation data---the RBLQNN is compared against mean-variance estimation networks (which presuppose that conditional distributions are Gaussian) and linear quantile regression (where linear conditional dependence relationships are assumed). The RBLQNN is versatile, issuing conditional probability estimates that faithfully describe the target variable in the broad class of datasets considered.

We evaluate the RBLQNN on three synthetic datasets with known a priori distributions, demonstrating several advantages over other approaches. The RBLQNN performs well when the mean-variance estimation network or linear quantile regression is deficient, such as when the conditional probability distributions are non-Gaussian or the response variable depends nonlinearly on the regressors. Evaluations of the RBLQNN against other quantile neural network techniques demonstrates advantages of our approach: convergence for our method appears to be more stable than the cumulative increment approach of \citeA{padilla2022quantile}, whereas the ReLU bias loss reduces degenerate probability distributions due to quantile crossings without significantly degrading performance. Evaluations of the convergence of the RBLQNN over a large range of hyperparameters suggests that the RBLQNN trains stably over a broad hyperparameter space relative to the mean-variance estimation networks. A caveat to these results is that further hyperparameter tuning or architectural changes (e.g., more layers, different activation function or optimizer) might alter the results. Yet, the modification to the loss offers at least a useful and rigorous alternative than more ad-hoc changes to the network.   

Comparing the CRPS of the RBLQNN predictions against linear quantile regression and the mean-variance estimation network predictions illustrates the relative importance of capturing nonlinearities or non-Gaussian distributions in the representation of uncertainties. For the GSOD daily temperature maxima dataset, most stations have lower sample-averaged CRPS with the RBLQNN than with linear quantile regression, and comparisons of bootstrapped CRPS averages indicate that these differences are statistically significant at many locations. The relative performance of the RBLQNN over linear quantile regression implies that capturing nonlinear functional dependencies between temperature and local pressure or geopotential heights is paramount to constraining temperature uncertainties. Differences in CRPS between the RBLQNN and the mean-variance estimation network are smaller and not statistically significant, suggesting that allowing for non-Gaussian conditional probabilities is of secondary importance.

In the context of daily temperature maxima, the relative validity of the Gaussian approximation in many situations may be related to the principle of maximum entropy \cite{sura2015perspectives, jaynes1957information}: namely, that the distribution which maximizes the information entropy---i.e., the least informative distribution---under a set of given constraints is that which is most probable. Under the very limited constraints of given mean and variance, the Gaussian distribution maximizes the differential entropy, and thus, without further information constraints, the Gaussian approximation is valid. It was noted that the CRPS for the RBLQNN often was higher than the mean-variance estimation networks when much of the variance was explained by the deterministic component (MSE network $R^2 \approx 1$) or when very little of the variance was explained by the deterministic component ($R^2 \approx 0$). In such cases, the Gaussian approximation may be relatively valid because the inputs are uninformative about the conditional distribution, and thus the distribution is relatively unconstrained. On the other hand, in regimes where the inputs are informative to predicting temperature, but much of the variance is left unexplained by the deterministic functional, additional information constraints may apply and the maximal entropy distribution may be more accurately described by non-Gaussian probability distributions.

While the RBLQNN does not have statistically significantly better CRPS than the mean-variance estimation networks on the GSOD dataset, it is possible that the RBLQNN does predict conditional distributions of temperature more skillfully, yet that the sample size is insufficient to discern this skill. The synthetic datasets in Section~\ref{sec:synthetic_results} demonstrate that even if the  RBLQNN clearly predicts the true distribution with greater accuracy (e.g., Fig.~\ref{fig:toy_data_quantile_mae}) a large number of samples may be needed to distinguish skill using sample-based metrics like the CRPS (e.g., Fig.~\ref{fig:toy_data_crps}). Station temperature observations with significantly more samples may more clearly reveal the skill of the RBLQNN, though it is difficult a priori to estimate how many samples are needed for non-Gaussian statistics to emerge.

In light of the meager sample sizes for the GSOD temperature datasets and challenges in identifying non-Gaussian conditional distributions, we also evaluated the performance of the RBLQNN on the TRMM precipitation dataset, in which the sample size was over ten times as large and the Gaussian approximation is clearly invalid. In this case, the RBLQNN clearly outperforms the linear quantile regression and mean-variance estimation network baselines. In principle, maximum-likelihood loss functions for other probability distributions can be used to estimate the parameters of different families of distributions. Since precipitation can take only nonnegative values, precipitation may be better modeled using distributions supported on the semi-infinite line, such as the exponential distribution or the Gamma distribution. Nevertheless, the RBLQNN is a simple approach for estimating conditional probabilities that does not require any assumptions about the parametric family of the underlying distribution.

\section*{Open Research Section}
The NOAA Global Surface Summary of the Day dataset is available at \url{https://www.ncei.noaa.gov/access/metadata/landing-page/bin/iso?id=gov.noaa.ncdc:C00516} \cite{noaa1999gsod}. The Tropical Rainfall Measuring Mission dataset is available at \url{https://gpm.nasa.gov/data/directory} \cite{kummerow2000status}. The ERA5 reanalysis dataset is available at \url{https://github.com/google-research/arco-era5} \cite{google2022era5}, courtesy of the Copernicus Climate Changes Service (C3S) Data Store \cite{hersbach2000decomposition}. The code used for data processing, model training, analysis, and visualization in this study and is archived at \url{https://doi.org/10.5281/zenodo.18355928} available under the MIT license at \url{https://github.com/andrewbrettin/rblqnn} \cite{brettin2025code}.

\acknowledgments
AB is supported by the VoLo Foundation. LZ received support through Schmidt Sciences, LLC, under the M$^2$LInES project. 
We thank Libby Barnes, Sara Shamekh, Carlos Fernandez-Granda, and Fabrizio Falasca for helpful discussions on this work.

%
%

\bibliography{bibliography}

\begin{thebibliography}{}

\bibitem [\protect \citeauthoryear {%
Ahmed%
\ \BBA {} Schumacher%
}{%
Ahmed%
\ \BBA {} Schumacher%
}{%
{\protect \APACyear {2015}}%
}]{%
ahmed2015convective}
\APACinsertmetastar {%
ahmed2015convective}%
\begin{APACrefauthors}%
Ahmed, F.%
\BCBT {}\ \BBA {} Schumacher, C.%
\end{APACrefauthors}%
\unskip\
\newblock
\APACrefYearMonthDay{2015}{}{}.
\newblock
{\BBOQ}\APACrefatitle {{Convective and stratiform components of the
  precipitation-moisture relationship}} {{Convective and stratiform components
  of the precipitation-moisture relationship}}.{\BBCQ}
\newblock
\APACjournalVolNumPages{{Geophysical Research Letters}}{42}{23}{10--453}.
\PrintBackRefs{\CurrentBib}

\bibitem [\protect \citeauthoryear {%
Ashkenazy%
\ \BBA {} Smith%
}{%
Ashkenazy%
\ \BBA {} Smith%
}{%
{\protect \APACyear {2024}}%
}]{%
ashkenazy2024data}
\APACinsertmetastar {%
ashkenazy2024data}%
\begin{APACrefauthors}%
Ashkenazy, Y.%
\BCBT {}\ \BBA {} Smith, N\BPBI R.%
\end{APACrefauthors}%
\unskip\
\newblock
\APACrefYearMonthDay{2024}{}{}.
\newblock
{\BBOQ}\APACrefatitle {{Data-driven analysis of annual rain distributions}}
  {{Data-driven analysis of annual rain distributions}}.{\BBCQ}
\newblock
\APACjournalVolNumPages{Physical Review Research}{6}{2}{023187}.
\PrintBackRefs{\CurrentBib}

\bibitem [\protect \citeauthoryear {%
Barnes%
\ \BBA {} Barnes%
}{%
Barnes%
\ \BBA {} Barnes%
}{%
{\protect \APACyear {2021}}%
}]{%
barnes2021controlled}
\APACinsertmetastar {%
barnes2021controlled}%
\begin{APACrefauthors}%
Barnes, E\BPBI A.%
\BCBT {}\ \BBA {} Barnes, R\BPBI J.%
\end{APACrefauthors}%
\unskip\
\newblock
\APACrefYearMonthDay{2021}{}{}.
\newblock
{\BBOQ}\APACrefatitle {{Controlled Abstention Neural Networks for Identifying
  Skillful Predictions for Regression Problems}} {{Controlled Abstention Neural
  Networks for Identifying Skillful Predictions for Regression
  Problems}}.{\BBCQ}
\newblock
\APACjournalVolNumPages{{Journal of Advances in Modeling Earth
  Systems}}{13}{12}{e2021MS002575}.
\PrintBackRefs{\CurrentBib}

\bibitem [\protect \citeauthoryear {%
Barnes%
, Barnes%
\BCBL {}\ \BBA {} DeMaria%
}{%
Barnes%
\ \protect \BOthers {.}}{%
{\protect \APACyear {2023}}%
}]{%
barnes2023sinh}
\APACinsertmetastar {%
barnes2023sinh}%
\begin{APACrefauthors}%
Barnes, E\BPBI A.%
, Barnes, R\BPBI J.%
\BCBL {}\ \BBA {} DeMaria, M.%
\end{APACrefauthors}%
\unskip\
\newblock
\APACrefYearMonthDay{2023}{}{}.
\newblock
{\BBOQ}\APACrefatitle {{Sinh-arcsinh-normal distributions to add uncertainty to
  neural network regression tasks: Applications to tropical cyclone intensity
  forecasts}} {{Sinh-arcsinh-normal distributions to add uncertainty to neural
  network regression tasks: Applications to tropical cyclone intensity
  forecasts}}.{\BBCQ}
\newblock
\APACjournalVolNumPages{{Environmental Data Science}}{2}{}{e15}.
\PrintBackRefs{\CurrentBib}

\bibitem [\protect \citeauthoryear {%
Beck%
\ \protect \BOthers {.}}{%
Beck%
\ \protect \BOthers {.}}{%
{\protect \APACyear {2020}}%
}]{%
beck2020ppdist}
\APACinsertmetastar {%
beck2020ppdist}%
\begin{APACrefauthors}%
Beck, H\BPBI E.%
, Westra, S.%
, Tan, J.%
, Pappenberger, F.%
, Huffman, G\BPBI J.%
, McVicar, T\BPBI R.%
\BDBL {}others%
\end{APACrefauthors}%
\unskip\
\newblock
\APACrefYearMonthDay{2020}{}{}.
\newblock
{\BBOQ}\APACrefatitle {{PPDIST, global 0.1${}^\circ$ daily and 3-hourly
  precipitation probability distribution climatologies for 1979--2018}}
  {{PPDIST, global 0.1${}^\circ$ daily and 3-hourly precipitation probability
  distribution climatologies for 1979--2018}}.{\BBCQ}
\newblock
\APACjournalVolNumPages{Scientific data}{7}{1}{302}.
\PrintBackRefs{\CurrentBib}

\bibitem [\protect \citeauthoryear {%
Belloni%
, Chernozhukov%
, Chetverikov%
\BCBL {}\ \BBA {} Fern{\'a}ndez-Val%
}{%
Belloni%
\ \protect \BOthers {.}}{%
{\protect \APACyear {2019}}%
}]{%
belloni2019conditional}
\APACinsertmetastar {%
belloni2019conditional}%
\begin{APACrefauthors}%
Belloni, A.%
, Chernozhukov, V.%
, Chetverikov, D.%
\BCBL {}\ \BBA {} Fern{\'a}ndez-Val, I.%
\end{APACrefauthors}%
\unskip\
\newblock
\APACrefYearMonthDay{2019}{}{}.
\newblock
{\BBOQ}\APACrefatitle {{Conditional quantile processes based on series or many
  regressors}} {{Conditional quantile processes based on series or many
  regressors}}.{\BBCQ}
\newblock
\APACjournalVolNumPages{{Journal of Econometrics}}{213}{1}{4--29}.
\PrintBackRefs{\CurrentBib}

\bibitem [\protect \citeauthoryear {%
Bjarke%
, Barsugli%
, Hoerling%
, Quan%
\BCBL {}\ \BBA {} Livneh%
}{%
Bjarke%
\ \protect \BOthers {.}}{%
{\protect \APACyear {2023}}%
}]{%
bjarke2023record}
\APACinsertmetastar {%
bjarke2023record}%
\begin{APACrefauthors}%
Bjarke, N.%
, Barsugli, J.%
, Hoerling, M.%
, Quan, X\BHBI W.%
\BCBL {}\ \BBA {} Livneh, B.%
\end{APACrefauthors}%
\unskip\
\newblock
\APACrefYearMonthDay{2023}{}{}.
\newblock
{\BBOQ}\APACrefatitle {{When record breaking heat waves should not surprise:
  skewness, heavy tails and implications for risk assessment}} {{When record
  breaking heat waves should not surprise: skewness, heavy tails and
  implications for risk assessment}}.{\BBCQ}
\newblock
\APACjournalVolNumPages{ESS Open Archive}{}{}{}.
\PrintBackRefs{\CurrentBib}

\bibitem [\protect \citeauthoryear {%
Bocquet%
, Pires%
\BCBL {}\ \BBA {} Wu%
}{%
Bocquet%
\ \protect \BOthers {.}}{%
{\protect \APACyear {2010}}%
}]{%
bocquet2010beyond}
\APACinsertmetastar {%
bocquet2010beyond}%
\begin{APACrefauthors}%
Bocquet, M.%
, Pires, C\BPBI A.%
\BCBL {}\ \BBA {} Wu, L.%
\end{APACrefauthors}%
\unskip\
\newblock
\APACrefYearMonthDay{2010}{}{}.
\newblock
{\BBOQ}\APACrefatitle {{Beyond Gaussian Statistical Modeling in Geophysical
  Data Assimilation}} {{Beyond Gaussian Statistical Modeling in Geophysical
  Data Assimilation}}.{\BBCQ}
\newblock
\APACjournalVolNumPages{Monthly Weather Review}{138}{8}{2997--3023}.
\PrintBackRefs{\CurrentBib}

\bibitem [\protect \citeauthoryear {%
Bourdin%
, Nipen%
\BCBL {}\ \BBA {} Stull%
}{%
Bourdin%
\ \protect \BOthers {.}}{%
{\protect \APACyear {2014}}%
}]{%
bourdin2014reliable}
\APACinsertmetastar {%
bourdin2014reliable}%
\begin{APACrefauthors}%
Bourdin, D\BPBI R.%
, Nipen, T\BPBI N.%
\BCBL {}\ \BBA {} Stull, R\BPBI B.%
\end{APACrefauthors}%
\unskip\
\newblock
\APACrefYearMonthDay{2014}{}{}.
\newblock
{\BBOQ}\APACrefatitle {{Reliable probabilistic forecasts from an ensemble
  reservoir inflow forecasting system}} {{Reliable probabilistic forecasts from
  an ensemble reservoir inflow forecasting system}}.{\BBCQ}
\newblock
\APACjournalVolNumPages{{Water Resources Research}}{50}{4}{3108--3130}.
\PrintBackRefs{\CurrentBib}

\bibitem [\protect \citeauthoryear {%
Bremnes%
}{%
Bremnes%
}{%
{\protect \APACyear {2020}}%
}]{%
bremnes2020ensemble}
\APACinsertmetastar {%
bremnes2020ensemble}%
\begin{APACrefauthors}%
Bremnes, J\BPBI B.%
\end{APACrefauthors}%
\unskip\
\newblock
\APACrefYearMonthDay{2020}{}{}.
\newblock
{\BBOQ}\APACrefatitle {{Ensemble Postprocessing Using Quantile Function
  Regression Based on Neural Networks and Bernstein Polynomials}} {{Ensemble
  Postprocessing Using Quantile Function Regression Based on Neural Networks
  and Bernstein Polynomials}}.{\BBCQ}
\newblock
\APACjournalVolNumPages{Monthly Weather Review}{148}{1}{403--414}.
\PrintBackRefs{\CurrentBib}

\bibitem [\protect \citeauthoryear {%
Brettin%
}{%
Brettin%
}{%
{\protect \APACyear {2026}}%
}]{%
brettin2025code}
\APACinsertmetastar {%
brettin2025code}%
\begin{APACrefauthors}%
Brettin, A.%
\end{APACrefauthors}%
\unskip\
\newblock
\APACrefYearMonthDay{2026}{01}{}.
\newblock
\APACrefbtitle {{Code for Brettin and Zanna (2025): Estimation of temperature
  and precipitation uncertainties using quantile neural networks}.} {{Code for
  Brettin and Zanna (2025): Estimation of temperature and precipitation
  uncertainties using quantile neural networks}.}
\newblock
\APACaddressPublisher{}{Zenodo}.
\newblock
\APACrefnote{\url{https://github.com/andrewbrettin/rblqnn}}
\newblock
\begin{APACrefDOI} \doi{10.5281/zenodo.18355928} \end{APACrefDOI}
\PrintBackRefs{\CurrentBib}

\bibitem [\protect \citeauthoryear {%
Br{\"o}cker%
}{%
Br{\"o}cker%
}{%
{\protect \APACyear {2012}}%
}]{%
brocker2012evaluating}
\APACinsertmetastar {%
brocker2012evaluating}%
\begin{APACrefauthors}%
Br{\"o}cker, J.%
\end{APACrefauthors}%
\unskip\
\newblock
\APACrefYearMonthDay{2012}{}{}.
\newblock
{\BBOQ}\APACrefatitle {{Evaluating raw ensembles with the continuous ranked
  probability score}} {{Evaluating raw ensembles with the continuous ranked
  probability score}}.{\BBCQ}
\newblock
\APACjournalVolNumPages{{Quarterly Journal of the Royal Meteorological
  Society}}{138}{667}{1611--1617}.
\PrintBackRefs{\CurrentBib}

\bibitem [\protect \citeauthoryear {%
Cannon%
}{%
Cannon%
}{%
{\protect \APACyear {2018}}%
}]{%
cannon2018non}
\APACinsertmetastar {%
cannon2018non}%
\begin{APACrefauthors}%
Cannon, A\BPBI J.%
\end{APACrefauthors}%
\unskip\
\newblock
\APACrefYearMonthDay{2018}{}{}.
\newblock
{\BBOQ}\APACrefatitle {{Non-crossing nonlinear regression quantiles by monotone
  composite quantile regression neural network, with application to rainfall
  extremes}} {{Non-crossing nonlinear regression quantiles by monotone
  composite quantile regression neural network, with application to rainfall
  extremes}}.{\BBCQ}
\newblock
\APACjournalVolNumPages{{Stochastic environmental research and risk
  assessment}}{32}{11}{3207--3225}.
\PrintBackRefs{\CurrentBib}

\bibitem [\protect \citeauthoryear {%
Carver%
\ \BBA {} Merose%
}{%
Carver%
\ \BBA {} Merose%
}{%
{\protect \APACyear {2023}}%
}]{%
google2022era5}
\APACinsertmetastar {%
google2022era5}%
\begin{APACrefauthors}%
Carver, R\BPBI W.%
\BCBT {}\ \BBA {} Merose, A.%
\end{APACrefauthors}%
\unskip\
\newblock
\APACrefYearMonthDay{2023}{}{}.
\newblock
{\BBOQ}\APACrefatitle {{ARCO-ERA5: An Analysis-Ready Cloud-Optimized Reanalysis
  Dataset}} {{ARCO-ERA5: An Analysis-Ready Cloud-Optimized Reanalysis
  Dataset}}.{\BBCQ}
\newblock
\BIn{} \APACrefbtitle {{22nd Conf. on AI for Env. Science}.} {{22nd Conf. on AI
  for Env. Science}.}
\newblock
\APACaddressPublisher{Denver, CO, USA}{{American Meteorological Society}}.
\newblock
\APACrefnote{\url{https://ams.confex.com/ams/103ANNUAL/meetingapp.cgi/Paper/415842}}
\PrintBackRefs{\CurrentBib}

\bibitem [\protect \citeauthoryear {%
Catalano%
, Loikith%
\BCBL {}\ \BBA {} Neelin%
}{%
Catalano%
\ \protect \BOthers {.}}{%
{\protect \APACyear {2021}}%
}]{%
catalano2021diagnosing}
\APACinsertmetastar {%
catalano2021diagnosing}%
\begin{APACrefauthors}%
Catalano, A.%
, Loikith, P.%
\BCBL {}\ \BBA {} Neelin, J.%
\end{APACrefauthors}%
\unskip\
\newblock
\APACrefYearMonthDay{2021}{}{}.
\newblock
{\BBOQ}\APACrefatitle {{Diagnosing Non-Gaussian Temperature Distribution Tails
  Using Back-Trajectory Analysis}} {{Diagnosing Non-Gaussian Temperature
  Distribution Tails Using Back-Trajectory Analysis}}.{\BBCQ}
\newblock
\APACjournalVolNumPages{Journal of Geophysical Research:
  Atmospheres}{126}{8}{e2020JD033726}.
\PrintBackRefs{\CurrentBib}

\bibitem [\protect \citeauthoryear {%
Cavanaugh%
\ \BBA {} Shen%
}{%
Cavanaugh%
\ \BBA {} Shen%
}{%
{\protect \APACyear {2014}}%
}]{%
cavanaugh2014northern}
\APACinsertmetastar {%
cavanaugh2014northern}%
\begin{APACrefauthors}%
Cavanaugh, N\BPBI R.%
\BCBT {}\ \BBA {} Shen, S\BPBI S.%
\end{APACrefauthors}%
\unskip\
\newblock
\APACrefYearMonthDay{2014}{}{}.
\newblock
{\BBOQ}\APACrefatitle {{Northern Hemisphere Climatology and Trends of
  Statistical Moments Documented from GHCN-Daily Surface Air Temperature
  Station Data from 1950 to 2010}} {{Northern Hemisphere Climatology and Trends
  of Statistical Moments Documented from GHCN-Daily Surface Air Temperature
  Station Data from 1950 to 2010}}.{\BBCQ}
\newblock
\APACjournalVolNumPages{Journal of climate}{27}{14}{5396--5410}.
\PrintBackRefs{\CurrentBib}

\bibitem [\protect \citeauthoryear {%
Chernozhukov%
, Fern{\'a}ndez-Val%
\BCBL {}\ \BBA {} Galichon%
}{%
Chernozhukov%
\ \protect \BOthers {.}}{%
{\protect \APACyear {2010}}%
}]{%
chernozhukov2010quantile}
\APACinsertmetastar {%
chernozhukov2010quantile}%
\begin{APACrefauthors}%
Chernozhukov, V.%
, Fern{\'a}ndez-Val, I.%
\BCBL {}\ \BBA {} Galichon, A.%
\end{APACrefauthors}%
\unskip\
\newblock
\APACrefYearMonthDay{2010}{}{}.
\newblock
{\BBOQ}\APACrefatitle {{Quantile and probability curves without crossing}}
  {{Quantile and probability curves without crossing}}.{\BBCQ}
\newblock
\APACjournalVolNumPages{{Econometrica}}{78}{3}{1093--1125}.
\PrintBackRefs{\CurrentBib}

\bibitem [\protect \citeauthoryear {%
Chronopoulos%
, Raftapostolos%
\BCBL {}\ \BBA {} Kapetanios%
}{%
Chronopoulos%
\ \protect \BOthers {.}}{%
{\protect \APACyear {2024}}%
}]{%
chronopoulos2024forecasting}
\APACinsertmetastar {%
chronopoulos2024forecasting}%
\begin{APACrefauthors}%
Chronopoulos, I.%
, Raftapostolos, A.%
\BCBL {}\ \BBA {} Kapetanios, G.%
\end{APACrefauthors}%
\unskip\
\newblock
\APACrefYearMonthDay{2024}{}{}.
\newblock
{\BBOQ}\APACrefatitle {{Forecasting Value-at-Risk Using Deep Neural Network
  Quantile Regression}} {{Forecasting Value-at-Risk Using Deep Neural Network
  Quantile Regression}}.{\BBCQ}
\newblock
\APACjournalVolNumPages{{Journal of Financial Econometrics}}{22}{3}{636--669}.
\PrintBackRefs{\CurrentBib}

\bibitem [\protect \citeauthoryear {%
Corsaro%
, Marino%
\BCBL {}\ \BBA {} Scognamiglio%
}{%
Corsaro%
\ \protect \BOthers {.}}{%
{\protect \APACyear {2024}}%
}]{%
corsaro2024quantile}
\APACinsertmetastar {%
corsaro2024quantile}%
\begin{APACrefauthors}%
Corsaro, S.%
, Marino, Z.%
\BCBL {}\ \BBA {} Scognamiglio, S.%
\end{APACrefauthors}%
\unskip\
\newblock
\APACrefYearMonthDay{2024}{}{}.
\newblock
{\BBOQ}\APACrefatitle {{Quantile mortality modelling of multiple populations
  via neural networks}} {{Quantile mortality modelling of multiple populations
  via neural networks}}.{\BBCQ}
\newblock
\APACjournalVolNumPages{Insurance: Mathematics and Economics}{116}{}{114--133}.
\PrintBackRefs{\CurrentBib}

\bibitem [\protect \citeauthoryear {%
Dawid%
}{%
Dawid%
}{%
{\protect \APACyear {1982}}%
}]{%
dawid1982well}
\APACinsertmetastar {%
dawid1982well}%
\begin{APACrefauthors}%
Dawid, A\BPBI P.%
\end{APACrefauthors}%
\unskip\
\newblock
\APACrefYearMonthDay{1982}{}{}.
\newblock
{\BBOQ}\APACrefatitle {{The Well-Calibrated Bayesian}} {{The Well-Calibrated
  Bayesian}}.{\BBCQ}
\newblock
\APACjournalVolNumPages{{Journal of the American Statistical
  Association}}{77}{379}{605--610}.
\PrintBackRefs{\CurrentBib}

\bibitem [\protect \citeauthoryear {%
Dawid%
}{%
Dawid%
}{%
{\protect \APACyear {1984}}%
}]{%
dawid1984present}
\APACinsertmetastar {%
dawid1984present}%
\begin{APACrefauthors}%
Dawid, A\BPBI P.%
\end{APACrefauthors}%
\unskip\
\newblock
\APACrefYearMonthDay{1984}{}{}.
\newblock
{\BBOQ}\APACrefatitle {{Statistical Theory: The Prequential Approach}}
  {{Statistical Theory: The Prequential Approach}}.{\BBCQ}
\newblock
\APACjournalVolNumPages{{Journal of the Royal Statistical Society, Series
  A}}{147}{2}{278--290}.
\PrintBackRefs{\CurrentBib}

\bibitem [\protect \citeauthoryear {%
DelSole%
\ \BBA {} Tippett%
}{%
DelSole%
\ \BBA {} Tippett%
}{%
{\protect \APACyear {2022}}%
}]{%
delsole2022statistical1}
\APACinsertmetastar {%
delsole2022statistical1}%
\begin{APACrefauthors}%
DelSole, T.%
\BCBT {}\ \BBA {} Tippett, M.%
\end{APACrefauthors}%
\unskip\
\newblock
\APACrefYearMonthDay{2022}{}{}.
\newblock
{\BBOQ}\APACrefatitle {Basic Concepts in Probability and Statistics} {Basic
  concepts in probability and statistics}.{\BBCQ}
\newblock
\BIn{} \APACrefbtitle {Statistical Methods for Climate Scientists} {Statistical
  methods for climate scientists}\ (\BPG~1–29).
\newblock
\APACaddressPublisher{}{Cambridge University Press}.
\PrintBackRefs{\CurrentBib}

\bibitem [\protect \citeauthoryear {%
Diffenbaugh%
\ \BBA {} Barnes%
}{%
Diffenbaugh%
\ \BBA {} Barnes%
}{%
{\protect \APACyear {2023}}%
}]{%
diffenbaugh2023data}
\APACinsertmetastar {%
diffenbaugh2023data}%
\begin{APACrefauthors}%
Diffenbaugh, N\BPBI S.%
\BCBT {}\ \BBA {} Barnes, E\BPBI A.%
\end{APACrefauthors}%
\unskip\
\newblock
\APACrefYearMonthDay{2023}{}{}.
\newblock
{\BBOQ}\APACrefatitle {{Data-driven predictions of the time remaining until
  critical global warming thresholds are reached}} {{Data-driven predictions of
  the time remaining until critical global warming thresholds are
  reached}}.{\BBCQ}
\newblock
\APACjournalVolNumPages{{Proceedings of the National Academy of
  Sciences}}{120}{6}{e2207183120}.
\PrintBackRefs{\CurrentBib}

\bibitem [\protect \citeauthoryear {%
Falasca%
\ \protect \BOthers {.}}{%
Falasca%
\ \protect \BOthers {.}}{%
{\protect \APACyear {2023}}%
}]{%
falasca2023exploring}
\APACinsertmetastar {%
falasca2023exploring}%
\begin{APACrefauthors}%
Falasca, F.%
, Brettin, A.%
, Zanna, L.%
, Griffies, S\BPBI M.%
, Yin, J.%
\BCBL {}\ \BBA {} Zhao, M.%
\end{APACrefauthors}%
\unskip\
\newblock
\APACrefYearMonthDay{2023}{}{}.
\newblock
{\BBOQ}\APACrefatitle {{Exploring the nonstationarity of coastal sea level
  probability distributions}} {{Exploring the nonstationarity of coastal sea
  level probability distributions}}.{\BBCQ}
\newblock
\APACjournalVolNumPages{{Environmental Data Science}}{2}{}{e16}.
\PrintBackRefs{\CurrentBib}

\bibitem [\protect \citeauthoryear {%
Fisher%
}{%
Fisher%
}{%
{\protect \APACyear {1970}}%
}]{%
fisher1970statistical}
\APACinsertmetastar {%
fisher1970statistical}%
\begin{APACrefauthors}%
Fisher, R\BPBI A.%
\end{APACrefauthors}%
\unskip\
\newblock
\APACrefYearMonthDay{1970}{}{}.
\newblock
{\BBOQ}\APACrefatitle {{Statistical methods for research workers}}
  {{Statistical methods for research workers}}.{\BBCQ}
\newblock
\BIn{} \APACrefbtitle {Breakthroughs in statistics: Methodology and
  distribution} {Breakthroughs in statistics: Methodology and distribution}\
  (\BPGS\ 66--70).
\newblock
\APACaddressPublisher{}{Springer}.
\PrintBackRefs{\CurrentBib}

\bibitem [\protect \citeauthoryear {%
Franzke%
, O'Kane%
, Berner%
, Williams%
\BCBL {}\ \BBA {} Lucarini%
}{%
Franzke%
\ \protect \BOthers {.}}{%
{\protect \APACyear {2015}}%
}]{%
franzke2015stochastic}
\APACinsertmetastar {%
franzke2015stochastic}%
\begin{APACrefauthors}%
Franzke, C\BPBI L.%
, O'Kane, T\BPBI J.%
, Berner, J.%
, Williams, P\BPBI D.%
\BCBL {}\ \BBA {} Lucarini, V.%
\end{APACrefauthors}%
\unskip\
\newblock
\APACrefYearMonthDay{2015}{}{}.
\newblock
{\BBOQ}\APACrefatitle {{Stochastic climate theory and modeling}} {{Stochastic
  climate theory and modeling}}.{\BBCQ}
\newblock
\APACjournalVolNumPages{Wiley Interdisciplinary Reviews: Climate
  Change}{6}{1}{63--78}.
\PrintBackRefs{\CurrentBib}

\bibitem [\protect \citeauthoryear {%
Garfinkel%
\ \BBA {} Harnik%
}{%
Garfinkel%
\ \BBA {} Harnik%
}{%
{\protect \APACyear {2017}}%
}]{%
garfinkel2017non}
\APACinsertmetastar {%
garfinkel2017non}%
\begin{APACrefauthors}%
Garfinkel, C\BPBI I.%
\BCBT {}\ \BBA {} Harnik, N.%
\end{APACrefauthors}%
\unskip\
\newblock
\APACrefYearMonthDay{2017}{}{}.
\newblock
{\BBOQ}\APACrefatitle {{The Non-Gaussianity and Spatial Asymmetry of
  Temperature Extremes Relative to the Storm Track: The Role of Horizontal
  Advection}} {{The Non-Gaussianity and Spatial Asymmetry of Temperature
  Extremes Relative to the Storm Track: The Role of Horizontal
  Advection}}.{\BBCQ}
\newblock
\APACjournalVolNumPages{Journal of Climate}{30}{2}{445--464}.
\PrintBackRefs{\CurrentBib}

\bibitem [\protect \citeauthoryear {%
Gneiting%
, Balabdaoui%
\BCBL {}\ \BBA {} Raftery%
}{%
Gneiting%
\ \protect \BOthers {.}}{%
{\protect \APACyear {2007}}%
}]{%
gneiting2007probabilistic}
\APACinsertmetastar {%
gneiting2007probabilistic}%
\begin{APACrefauthors}%
Gneiting, T.%
, Balabdaoui, F.%
\BCBL {}\ \BBA {} Raftery, A\BPBI E.%
\end{APACrefauthors}%
\unskip\
\newblock
\APACrefYearMonthDay{2007}{}{}.
\newblock
{\BBOQ}\APACrefatitle {{Probabilistic forecasts, calibration and sharpness}}
  {{Probabilistic forecasts, calibration and sharpness}}.{\BBCQ}
\newblock
\APACjournalVolNumPages{{Journal of the Royal Statistical Society Series B:
  Statistical Methodology}}{69}{2}{243--268}.
\PrintBackRefs{\CurrentBib}

\bibitem [\protect \citeauthoryear {%
Gneiting%
\ \BBA {} Raftery%
}{%
Gneiting%
\ \BBA {} Raftery%
}{%
{\protect \APACyear {2007}}%
}]{%
gneiting2007strictly}
\APACinsertmetastar {%
gneiting2007strictly}%
\begin{APACrefauthors}%
Gneiting, T.%
\BCBT {}\ \BBA {} Raftery, A\BPBI E.%
\end{APACrefauthors}%
\unskip\
\newblock
\APACrefYearMonthDay{2007}{}{}.
\newblock
{\BBOQ}\APACrefatitle {{Strictly proper scoring rules, prediction, and
  estimation}} {{Strictly proper scoring rules, prediction, and
  estimation}}.{\BBCQ}
\newblock
\APACjournalVolNumPages{{Journal of Atmospheric Science}}{102}{477}{359--378}.
\PrintBackRefs{\CurrentBib}

\bibitem [\protect \citeauthoryear {%
Godbole%
, Dahl%
, Gilmer%
, Shallue%
\BCBL {}\ \BBA {} Nado%
}{%
Godbole%
\ \protect \BOthers {.}}{%
{\protect \APACyear {2023}}%
}]{%
godbole2023tuning}
\APACinsertmetastar {%
godbole2023tuning}%
\begin{APACrefauthors}%
Godbole, V.%
, Dahl, G\BPBI E.%
, Gilmer, J.%
, Shallue, C\BPBI J.%
\BCBL {}\ \BBA {} Nado, Z.%
\end{APACrefauthors}%
\unskip\
\newblock
\APACrefYearMonthDay{2023}{}{}.
\newblock
\APACrefbtitle {{Deep Learning Tuning Playbook}.} {{Deep Learning Tuning
  Playbook}.}
\newblock
\APACrefnote{Version 1.0,
  \url{http://github.com/google-research/tuning_playbook}}
\PrintBackRefs{\CurrentBib}

\bibitem [\protect \citeauthoryear {%
Goodfellow%
, Bengio%
\BCBL {}\ \BBA {} Courville%
}{%
Goodfellow%
\ \protect \BOthers {.}}{%
{\protect \APACyear {2016}}%
}]{%
goodfellow2016deep}
\APACinsertmetastar {%
goodfellow2016deep}%
\begin{APACrefauthors}%
Goodfellow, I.%
, Bengio, Y.%
\BCBL {}\ \BBA {} Courville, A.%
\end{APACrefauthors}%
\unskip\
\newblock
\APACrefYearMonthDay{2016}{}{}.
\newblock
{\BBOQ}\APACrefatitle {Deep Learning} {Deep learning}.{\BBCQ}
\newblock
\BIn{} (\BCHAP\ Chapter 8: Optimization for Training Deep Models).
\newblock
\APACaddressPublisher{}{MIT Press}.
\newblock
\APACrefnote{\url{http://www.deeplearningbook.org}}
\PrintBackRefs{\CurrentBib}

\bibitem [\protect \citeauthoryear {%
Gordon%
\ \BBA {} Barnes%
}{%
Gordon%
\ \BBA {} Barnes%
}{%
{\protect \APACyear {2022}}%
}]{%
gordon2022incorporating}
\APACinsertmetastar {%
gordon2022incorporating}%
\begin{APACrefauthors}%
Gordon, E\BPBI M.%
\BCBT {}\ \BBA {} Barnes, E\BPBI A.%
\end{APACrefauthors}%
\unskip\
\newblock
\APACrefYearMonthDay{2022}{}{}.
\newblock
{\BBOQ}\APACrefatitle {{Incorporating Uncertainty Into a Regression Neural
  Network Enables Identification of Decadal State-Dependent Predictability in
  CESM2}} {{Incorporating Uncertainty Into a Regression Neural Network Enables
  Identification of Decadal State-Dependent Predictability in CESM2}}.{\BBCQ}
\newblock
\APACjournalVolNumPages{{Geophysical Research Letters}}{49}{15}{e2022GL098635}.
\PrintBackRefs{\CurrentBib}

\bibitem [\protect \citeauthoryear {%
Guillaumin%
\ \BBA {} Zanna%
}{%
Guillaumin%
\ \BBA {} Zanna%
}{%
{\protect \APACyear {2021}}%
}]{%
guillaumin2021stochastic}
\APACinsertmetastar {%
guillaumin2021stochastic}%
\begin{APACrefauthors}%
Guillaumin, A\BPBI P.%
\BCBT {}\ \BBA {} Zanna, L.%
\end{APACrefauthors}%
\unskip\
\newblock
\APACrefYearMonthDay{2021}{}{}.
\newblock
{\BBOQ}\APACrefatitle {{Stochastic-Deep Learning Parameterization of Ocean
  Momentum Forcing}} {{Stochastic-Deep Learning Parameterization of Ocean
  Momentum Forcing}}.{\BBCQ}
\newblock
\APACjournalVolNumPages{{Journal of Advances in Modeling Earth
  Systems}}{13}{9}{e2021MS002534}.
\PrintBackRefs{\CurrentBib}

\bibitem [\protect \citeauthoryear {%
Gupta%
, Mastrantonas%
, Masoller%
\BCBL {}\ \BBA {} Kurths%
}{%
Gupta%
\ \protect \BOthers {.}}{%
{\protect \APACyear {2022}}%
}]{%
gupta2022perspectives}
\APACinsertmetastar {%
gupta2022perspectives}%
\begin{APACrefauthors}%
Gupta, S.%
, Mastrantonas, N.%
, Masoller, C.%
\BCBL {}\ \BBA {} Kurths, J.%
\end{APACrefauthors}%
\unskip\
\newblock
\APACrefYearMonthDay{2022}{}{}.
\newblock
{\BBOQ}\APACrefatitle {{Perspectives on the importance of complex systems in
  understanding our climate and climate change: The Nobel Prize in Physics
  2021}} {{Perspectives on the importance of complex systems in understanding
  our climate and climate change: The Nobel Prize in Physics 2021}}.{\BBCQ}
\newblock
\APACjournalVolNumPages{{Chaos: An Interdisciplinary Journal of Nonlinear
  Science}}{32}{5}{}.
\PrintBackRefs{\CurrentBib}

\bibitem [\protect \citeauthoryear {%
Hassanzadeh%
\ \BBA {} Kuang%
}{%
Hassanzadeh%
\ \BBA {} Kuang%
}{%
{\protect \APACyear {2015}}%
}]{%
hassanzadeh2015blocking}
\APACinsertmetastar {%
hassanzadeh2015blocking}%
\begin{APACrefauthors}%
Hassanzadeh, P.%
\BCBT {}\ \BBA {} Kuang, Z.%
\end{APACrefauthors}%
\unskip\
\newblock
\APACrefYearMonthDay{2015}{}{}.
\newblock
{\BBOQ}\APACrefatitle {{Blocking variability: Arctic Amplification versus
  Arctic Oscillation}} {{Blocking variability: Arctic Amplification versus
  Arctic Oscillation}}.{\BBCQ}
\newblock
\APACjournalVolNumPages{Geophysical Research Letters}{42}{20}{8586--8595}.
\PrintBackRefs{\CurrentBib}

\bibitem [\protect \citeauthoryear {%
Haynes%
, Lagerquist%
, McGraw%
, Musgrave%
\BCBL {}\ \BBA {} Ebert-Uphoff%
}{%
Haynes%
\ \protect \BOthers {.}}{%
{\protect \APACyear {2023}}%
}]{%
haynes2023creating}
\APACinsertmetastar {%
haynes2023creating}%
\begin{APACrefauthors}%
Haynes, K.%
, Lagerquist, R.%
, McGraw, M.%
, Musgrave, K.%
\BCBL {}\ \BBA {} Ebert-Uphoff, I.%
\end{APACrefauthors}%
\unskip\
\newblock
\APACrefYearMonthDay{2023}{}{}.
\newblock
{\BBOQ}\APACrefatitle {{Creating and Evaluating Uncertainty Estimates with
  Neural Networks for Environmental-Science Applications}} {{Creating and
  Evaluating Uncertainty Estimates with Neural Networks for
  Environmental-Science Applications}}.{\BBCQ}
\newblock
\APACjournalVolNumPages{Artificial Intelligence for the Earth
  Systems}{2}{2}{220061}.
\PrintBackRefs{\CurrentBib}

\bibitem [\protect \citeauthoryear {%
Hersbach%
}{%
Hersbach%
}{%
{\protect \APACyear {2000}}%
}]{%
hersbach2000decomposition}
\APACinsertmetastar {%
hersbach2000decomposition}%
\begin{APACrefauthors}%
Hersbach, H.%
\end{APACrefauthors}%
\unskip\
\newblock
\APACrefYearMonthDay{2000}{}{}.
\newblock
{\BBOQ}\APACrefatitle {{Decomposition of the Continuous Ranked Probability
  Score for Ensemble Prediction Systems}} {{Decomposition of the Continuous
  Ranked Probability Score for Ensemble Prediction Systems}}.{\BBCQ}
\newblock
\APACjournalVolNumPages{{Weather and Forecasting}}{15}{5}{559--570}.
\PrintBackRefs{\CurrentBib}

\bibitem [\protect \citeauthoryear {%
Hersbach%
\ \protect \BOthers {.}}{%
Hersbach%
\ \protect \BOthers {.}}{%
{\protect \APACyear {2020}}%
}]{%
hersbach2020era5}
\APACinsertmetastar {%
hersbach2020era5}%
\begin{APACrefauthors}%
Hersbach, H.%
, Bell, B.%
, Berrisford, P.%
, Hirahara, S.%
, Hor{\'a}nyi, A.%
, Mu{\~n}oz-Sabater, J.%
\BDBL {}others%
\end{APACrefauthors}%
\unskip\
\newblock
\APACrefYearMonthDay{2020}{}{}.
\newblock
{\BBOQ}\APACrefatitle {{The ERA5 global reanalysis}} {{The ERA5 global
  reanalysis}}.{\BBCQ}
\newblock
\APACjournalVolNumPages{{Quarterly Journal of the Royal Meteorological
  Society}}{146}{730}{1999--2049}.
\PrintBackRefs{\CurrentBib}

\bibitem [\protect \citeauthoryear {%
Hu%
\ \BBA {} Pierrehumbert%
}{%
Hu%
\ \BBA {} Pierrehumbert%
}{%
{\protect \APACyear {2002}}%
}]{%
hu2002advection}
\APACinsertmetastar {%
hu2002advection}%
\begin{APACrefauthors}%
Hu, Y.%
\BCBT {}\ \BBA {} Pierrehumbert, R\BPBI T.%
\end{APACrefauthors}%
\unskip\
\newblock
\APACrefYearMonthDay{2002}{}{}.
\newblock
{\BBOQ}\APACrefatitle {{The Advection–Diffusion Problem for Stratospheric
  Flow. Part II: Probability Distribution Function of Tracer Gradients}} {{The
  Advection–Diffusion Problem for Stratospheric Flow. Part II: Probability
  Distribution Function of Tracer Gradients}}.{\BBCQ}
\newblock
\APACjournalVolNumPages{Journal of the atmospheric
  sciences}{59}{19}{2830--2845}.
\PrintBackRefs{\CurrentBib}

\bibitem [\protect \citeauthoryear {%
Hyv{\"a}rinen%
\ \BBA {} Oja%
}{%
Hyv{\"a}rinen%
\ \BBA {} Oja%
}{%
{\protect \APACyear {2000}}%
}]{%
hyvarinen2000independent}
\APACinsertmetastar {%
hyvarinen2000independent}%
\begin{APACrefauthors}%
Hyv{\"a}rinen, A.%
\BCBT {}\ \BBA {} Oja, E.%
\end{APACrefauthors}%
\unskip\
\newblock
\APACrefYearMonthDay{2000}{}{}.
\newblock
{\BBOQ}\APACrefatitle {{Independent component analysis: algorithms and
  applications}} {{Independent component analysis: algorithms and
  applications}}.{\BBCQ}
\newblock
\APACjournalVolNumPages{Neural networks}{13}{4-5}{411--430}.
\PrintBackRefs{\CurrentBib}

\bibitem [\protect \citeauthoryear {%
{IPCC}%
}{%
{IPCC}%
}{%
{\protect \APACyear {2021}}%
}]{%
ipcc2022spm}
\APACinsertmetastar {%
ipcc2022spm}%
\begin{APACrefauthors}%
{IPCC}.%
\end{APACrefauthors}%
\unskip\
\newblock
\APACrefYearMonthDay{2021}{}{}.
\newblock
\APACrefbtitle {{Climate Change 2021: The Physical Science Basis. Contribution
  of Working Group I to the Sixth Assessment Report of the Intergovernmental
  Panel on Climate Change}} {{Climate Change 2021: The Physical Science Basis.
  Contribution of Working Group I to the Sixth Assessment Report of the
  Intergovernmental Panel on Climate Change}}\ \APACbVolEdTR{}{\BTR{}}.
\newblock
\APACaddressInstitution{{Cambridge, United Kingdom and New York, NY,
  USA}}{{Intergovernmental Panel on Climate Change}}.
\PrintBackRefs{\CurrentBib}

\bibitem [\protect \citeauthoryear {%
Jaynes%
}{%
Jaynes%
}{%
{\protect \APACyear {1957}}%
}]{%
jaynes1957information}
\APACinsertmetastar {%
jaynes1957information}%
\begin{APACrefauthors}%
Jaynes, E\BPBI T.%
\end{APACrefauthors}%
\unskip\
\newblock
\APACrefYearMonthDay{1957}{}{}.
\newblock
{\BBOQ}\APACrefatitle {{Information Theory and Statistical Mechanics}}
  {{Information Theory and Statistical Mechanics}}.{\BBCQ}
\newblock
\APACjournalVolNumPages{Physical Review}{106}{4}{620}.
\PrintBackRefs{\CurrentBib}

\bibitem [\protect \citeauthoryear {%
Jiang%
, Jiang%
\BCBL {}\ \BBA {} Song%
}{%
Jiang%
\ \protect \BOthers {.}}{%
{\protect \APACyear {2012}}%
}]{%
jiang2012oracle}
\APACinsertmetastar {%
jiang2012oracle}%
\begin{APACrefauthors}%
Jiang, X.%
, Jiang, J.%
\BCBL {}\ \BBA {} Song, X.%
\end{APACrefauthors}%
\unskip\
\newblock
\APACrefYearMonthDay{2012}{}{}.
\newblock
{\BBOQ}\APACrefatitle {{Oracle model selection for nonlinear models based on
  weighted composite quantile regression}} {{Oracle model selection for
  nonlinear models based on weighted composite quantile regression}}.{\BBCQ}
\newblock
\APACjournalVolNumPages{{Statistica Sinica}}{}{}{1479--1506}.
\PrintBackRefs{\CurrentBib}

\bibitem [\protect \citeauthoryear {%
Jones%
\ \BBA {} Pewsey%
}{%
Jones%
\ \BBA {} Pewsey%
}{%
{\protect \APACyear {2019}}%
}]{%
jones2019sinh}
\APACinsertmetastar {%
jones2019sinh}%
\begin{APACrefauthors}%
Jones, C.%
\BCBT {}\ \BBA {} Pewsey, A.%
\end{APACrefauthors}%
\unskip\
\newblock
\APACrefYearMonthDay{2019}{}{}.
\newblock
{\BBOQ}\APACrefatitle {{The sinh-arcsinh normal distribution}} {{The
  sinh-arcsinh normal distribution}}.{\BBCQ}
\newblock
\APACjournalVolNumPages{{Significance}}{16}{2}{6--7}.
\PrintBackRefs{\CurrentBib}

\bibitem [\protect \citeauthoryear {%
Keilbar%
\ \BBA {} Wang%
}{%
Keilbar%
\ \BBA {} Wang%
}{%
{\protect \APACyear {2022}}%
}]{%
keilbar2022modelling}
\APACinsertmetastar {%
keilbar2022modelling}%
\begin{APACrefauthors}%
Keilbar, G.%
\BCBT {}\ \BBA {} Wang, W.%
\end{APACrefauthors}%
\unskip\
\newblock
\APACrefYearMonthDay{2022}{}{}.
\newblock
{\BBOQ}\APACrefatitle {{Modelling systemic risk using neural network quantile
  regression}} {{Modelling systemic risk using neural network quantile
  regression}}.{\BBCQ}
\newblock
\APACjournalVolNumPages{{Empirical Economics}}{62}{1}{93--118}.
\PrintBackRefs{\CurrentBib}

\bibitem [\protect \citeauthoryear {%
Kimura%
\ \BBA {} Kraichnan%
}{%
Kimura%
\ \BBA {} Kraichnan%
}{%
{\protect \APACyear {1993}}%
}]{%
kimura1993statistics}
\APACinsertmetastar {%
kimura1993statistics}%
\begin{APACrefauthors}%
Kimura, Y.%
\BCBT {}\ \BBA {} Kraichnan, R\BPBI H.%
\end{APACrefauthors}%
\unskip\
\newblock
\APACrefYearMonthDay{1993}{}{}.
\newblock
{\BBOQ}\APACrefatitle {{Statistics of an advected passive scalar}} {{Statistics
  of an advected passive scalar}}.{\BBCQ}
\newblock
\APACjournalVolNumPages{Physics of Fluids A: Fluid Dynamics}{5}{9}{2264--2277}.
\PrintBackRefs{\CurrentBib}

\bibitem [\protect \citeauthoryear {%
Kingma%
\ \BBA {} Ba%
}{%
Kingma%
\ \BBA {} Ba%
}{%
{\protect \APACyear {2014}}%
}]{%
kingma2014adam}
\APACinsertmetastar {%
kingma2014adam}%
\begin{APACrefauthors}%
Kingma, D\BPBI P.%
\BCBT {}\ \BBA {} Ba, J.%
\end{APACrefauthors}%
\unskip\
\newblock
\APACrefYearMonthDay{2014}{}{}.
\newblock
{\BBOQ}\APACrefatitle {{Adam: A method for stochastic optimization}} {{Adam: A
  method for stochastic optimization}}.{\BBCQ}
\newblock
\APACjournalVolNumPages{{arXiv preprint arXiv:1412.6980}}{}{}{}.
\newblock
\begin{APACrefDOI} \doi{10.48550/arXiv.1412.6980} \end{APACrefDOI}
\PrintBackRefs{\CurrentBib}

\bibitem [\protect \citeauthoryear {%
Koenker%
}{%
Koenker%
}{%
{\protect \APACyear {2005}}%
}]{%
koenker2005quantile}
\APACinsertmetastar {%
koenker2005quantile}%
\begin{APACrefauthors}%
Koenker, R.%
\end{APACrefauthors}%
\unskip\
\newblock
\APACrefYear{2005}.
\newblock
\APACrefbtitle {{Quantile regression}} {{Quantile regression}}\ (\BVOL~38).
\newblock
\APACaddressPublisher{}{Cambridge university press}.
\PrintBackRefs{\CurrentBib}

\bibitem [\protect \citeauthoryear {%
Koenker%
\ \BBA {} Bassett~Jr%
}{%
Koenker%
\ \BBA {} Bassett~Jr%
}{%
{\protect \APACyear {1978}}%
}]{%
koenker1978regression}
\APACinsertmetastar {%
koenker1978regression}%
\begin{APACrefauthors}%
Koenker, R.%
\BCBT {}\ \BBA {} Bassett~Jr, G.%
\end{APACrefauthors}%
\unskip\
\newblock
\APACrefYearMonthDay{1978}{}{}.
\newblock
{\BBOQ}\APACrefatitle {{Regression quantiles}} {{Regression quantiles}}.{\BBCQ}
\newblock
\APACjournalVolNumPages{{Econometrica}}{}{}{33--50}.
\PrintBackRefs{\CurrentBib}

\bibitem [\protect \citeauthoryear {%
Kopp%
\ \protect \BOthers {.}}{%
Kopp%
\ \protect \BOthers {.}}{%
{\protect \APACyear {2014}}%
}]{%
kopp2014probabilistic}
\APACinsertmetastar {%
kopp2014probabilistic}%
\begin{APACrefauthors}%
Kopp, R\BPBI E.%
, Horton, R\BPBI M.%
, Little, C\BPBI M.%
, Mitrovica, J\BPBI X.%
, Oppenheimer, M.%
, Rasmussen, D.%
\BDBL {}Tebaldi, C.%
\end{APACrefauthors}%
\unskip\
\newblock
\APACrefYearMonthDay{2014}{}{}.
\newblock
{\BBOQ}\APACrefatitle {{Probabilistic 21st and 22nd century sea-level
  projections at a global network of tide-gauge sites}} {{Probabilistic 21st
  and 22nd century sea-level projections at a global network of tide-gauge
  sites}}.{\BBCQ}
\newblock
\APACjournalVolNumPages{{Earth's future}}{2}{8}{383--406}.
\PrintBackRefs{\CurrentBib}

\bibitem [\protect \citeauthoryear {%
Kummerow%
\ \protect \BOthers {.}}{%
Kummerow%
\ \protect \BOthers {.}}{%
{\protect \APACyear {2000}}%
}]{%
kummerow2000status}
\APACinsertmetastar {%
kummerow2000status}%
\begin{APACrefauthors}%
Kummerow, C.%
, Simpson, J.%
, Thiele, O.%
, Barnes, W.%
, Chang, A.%
, Stocker, E.%
\BDBL {}others%
\end{APACrefauthors}%
\unskip\
\newblock
\APACrefYearMonthDay{2000}{}{}.
\newblock
{\BBOQ}\APACrefatitle {{The status of the Tropical Rainfall Measuring Mission
  (TRMM) after two years in orbit}} {{The status of the Tropical Rainfall
  Measuring Mission (TRMM) after two years in orbit}}.{\BBCQ}
\newblock
\APACjournalVolNumPages{{Journal of Applied Meteorology}}{39}{12}{1965--1982}.
\PrintBackRefs{\CurrentBib}

\bibitem [\protect \citeauthoryear {%
Laio%
\ \BBA {} Tamea%
}{%
Laio%
\ \BBA {} Tamea%
}{%
{\protect \APACyear {2007}}%
}]{%
laio2007verification}
\APACinsertmetastar {%
laio2007verification}%
\begin{APACrefauthors}%
Laio, F.%
\BCBT {}\ \BBA {} Tamea, S.%
\end{APACrefauthors}%
\unskip\
\newblock
\APACrefYearMonthDay{2007}{}{}.
\newblock
{\BBOQ}\APACrefatitle {{Verification tools for probabilistic forecasts of
  continuous hydrological variables}} {{Verification tools for probabilistic
  forecasts of continuous hydrological variables}}.{\BBCQ}
\newblock
\APACjournalVolNumPages{{Hydrology and Earth System
  Sciences}}{11}{4}{1267--1277}.
\PrintBackRefs{\CurrentBib}

\bibitem [\protect \citeauthoryear {%
Landau%
\ \BBA {} Lifshitz%
}{%
Landau%
\ \BBA {} Lifshitz%
}{%
{\protect \APACyear {2013}}%
}]{%
landau2013statistical}
\APACinsertmetastar {%
landau2013statistical}%
\begin{APACrefauthors}%
Landau, L\BPBI D.%
\BCBT {}\ \BBA {} Lifshitz, E\BPBI M.%
\end{APACrefauthors}%
\unskip\
\newblock
\APACrefYear{2013}.
\newblock
\APACrefbtitle {{Statistical Physics: Volume 5}} {{Statistical Physics: Volume
  5}}\ (\BVOL~5).
\newblock
\APACaddressPublisher{}{Elsevier}.
\PrintBackRefs{\CurrentBib}

\bibitem [\protect \citeauthoryear {%
Li%
, Wang%
, Cao%
, Zong%
\BCBL {}\ \BBA {} Chai%
}{%
Li%
\ \protect \BOthers {.}}{%
{\protect \APACyear {2023}}%
}]{%
li2023determining}
\APACinsertmetastar {%
li2023determining}%
\begin{APACrefauthors}%
Li, M.%
, Wang, G.%
, Cao, F.%
, Zong, S.%
\BCBL {}\ \BBA {} Chai, X.%
\end{APACrefauthors}%
\unskip\
\newblock
\APACrefYearMonthDay{2023}{}{}.
\newblock
{\BBOQ}\APACrefatitle {{Determining optimal probability distributions for
  gridded precipitation data based on L-moments}} {{Determining optimal
  probability distributions for gridded precipitation data based on
  L-moments}}.{\BBCQ}
\newblock
\APACjournalVolNumPages{Science of The Total Environment}{882}{}{163528}.
\PrintBackRefs{\CurrentBib}

\bibitem [\protect \citeauthoryear {%
Linz%
, Chen%
\BCBL {}\ \BBA {} Hu%
}{%
Linz%
\ \protect \BOthers {.}}{%
{\protect \APACyear {2018}}%
}]{%
linz2018large}
\APACinsertmetastar {%
linz2018large}%
\begin{APACrefauthors}%
Linz, M.%
, Chen, G.%
\BCBL {}\ \BBA {} Hu, Z.%
\end{APACrefauthors}%
\unskip\
\newblock
\APACrefYearMonthDay{2018}{}{}.
\newblock
{\BBOQ}\APACrefatitle {{Large-Scale Atmospheric Control on Non-Gaussian Tails
  of Midlatitude Temperature Distributions}} {{Large-Scale Atmospheric Control
  on Non-Gaussian Tails of Midlatitude Temperature Distributions}}.{\BBCQ}
\newblock
\APACjournalVolNumPages{Geophysical Research Letters}{45}{17}{9141--9149}.
\PrintBackRefs{\CurrentBib}

\bibitem [\protect \citeauthoryear {%
Loikith%
\ \BBA {} Neelin%
}{%
Loikith%
\ \BBA {} Neelin%
}{%
{\protect \APACyear {2015}}%
}]{%
loikith2015short}
\APACinsertmetastar {%
loikith2015short}%
\begin{APACrefauthors}%
Loikith, P\BPBI C.%
\BCBT {}\ \BBA {} Neelin, J\BPBI D.%
\end{APACrefauthors}%
\unskip\
\newblock
\APACrefYearMonthDay{2015}{}{}.
\newblock
{\BBOQ}\APACrefatitle {{Short-tailed temperature distributions over North
  America and implications for future changes in extremes}} {{Short-tailed
  temperature distributions over North America and implications for future
  changes in extremes}}.{\BBCQ}
\newblock
\APACjournalVolNumPages{Geophysical Research Letters}{42}{20}{8577--8585}.
\PrintBackRefs{\CurrentBib}

\bibitem [\protect \citeauthoryear {%
Loikith%
\ \BBA {} Neelin%
}{%
Loikith%
\ \BBA {} Neelin%
}{%
{\protect \APACyear {2019}}%
}]{%
loikith2019non}
\APACinsertmetastar {%
loikith2019non}%
\begin{APACrefauthors}%
Loikith, P\BPBI C.%
\BCBT {}\ \BBA {} Neelin, J\BPBI D.%
\end{APACrefauthors}%
\unskip\
\newblock
\APACrefYearMonthDay{2019}{}{}.
\newblock
{\BBOQ}\APACrefatitle {{Non-Gaussian Cold-Side Temperature Distribution Tails
  and Associated Synoptic Meteorology}} {{Non-Gaussian Cold-Side Temperature
  Distribution Tails and Associated Synoptic Meteorology}}.{\BBCQ}
\newblock
\APACjournalVolNumPages{Journal of Climate}{32}{23}{8399--8414}.
\PrintBackRefs{\CurrentBib}

\bibitem [\protect \citeauthoryear {%
Lott%
, Baldwin%
\BCBL {}\ \BBA {} Jones%
}{%
Lott%
\ \protect \BOthers {.}}{%
{\protect \APACyear {2001}}%
}]{%
lott2001isd}
\APACinsertmetastar {%
lott2001isd}%
\begin{APACrefauthors}%
Lott, N.%
, Baldwin, R.%
\BCBL {}\ \BBA {} Jones, P.%
\end{APACrefauthors}%
\unskip\
\newblock
\APACrefYearMonthDay{2001}{}{}.
\newblock
\APACrefbtitle {{The FCC Integrated Surface Hourly Database, A New Resource of
  Global Climate Data}} {{The FCC Integrated Surface Hourly Database, A New
  Resource of Global Climate Data}}\ \APACbVolEdTR{}{\BTR{}}.
\newblock
\APACaddressInstitution{Asheville, NC 28801-5696}{National Climatic Data
  Center}.
\newblock
\APAChowpublished
  {\url{https://repository.library.noaa.gov/view/noaa/13826/noaa_13826_DS1.pdf}}.
\PrintBackRefs{\CurrentBib}

\bibitem [\protect \citeauthoryear {%
Majda%
\ \BBA {} Wang%
}{%
Majda%
\ \BBA {} Wang%
}{%
{\protect \APACyear {2006}}%
}]{%
majda2006nonlinear}
\APACinsertmetastar {%
majda2006nonlinear}%
\begin{APACrefauthors}%
Majda, A.%
\BCBT {}\ \BBA {} Wang, X.%
\end{APACrefauthors}%
\unskip\
\newblock
\APACrefYear{2006}.
\newblock
\APACrefbtitle {{Nonlinear Dynamics and Statistical Theories for Basic
  Geophysical Flows}} {{Nonlinear Dynamics and Statistical Theories for Basic
  Geophysical Flows}}.
\newblock
\APACaddressPublisher{}{Cambridge University Press}.
\PrintBackRefs{\CurrentBib}

\bibitem [\protect \citeauthoryear {%
Martinez-Villalobos%
\ \BBA {} Neelin%
}{%
Martinez-Villalobos%
\ \BBA {} Neelin%
}{%
{\protect \APACyear {2019}}%
}]{%
martinez2019precipitation}
\APACinsertmetastar {%
martinez2019precipitation}%
\begin{APACrefauthors}%
Martinez-Villalobos, C.%
\BCBT {}\ \BBA {} Neelin, J\BPBI D.%
\end{APACrefauthors}%
\unskip\
\newblock
\APACrefYearMonthDay{2019}{}{}.
\newblock
{\BBOQ}\APACrefatitle {{Why Do Precipitation Intensities Tend to Follow Gamma
  Distributions?}} {{Why Do Precipitation Intensities Tend to Follow Gamma
  Distributions?}}{\BBCQ}
\newblock
\APACjournalVolNumPages{Journal of the Atmospheric
  Sciences}{76}{11}{3611--3631}.
\PrintBackRefs{\CurrentBib}

\bibitem [\protect \citeauthoryear {%
Matheson%
\ \BBA {} Winkler%
}{%
Matheson%
\ \BBA {} Winkler%
}{%
{\protect \APACyear {1976}}%
}]{%
matheson1976scoring}
\APACinsertmetastar {%
matheson1976scoring}%
\begin{APACrefauthors}%
Matheson, J\BPBI E.%
\BCBT {}\ \BBA {} Winkler, R\BPBI L.%
\end{APACrefauthors}%
\unskip\
\newblock
\APACrefYearMonthDay{1976}{}{}.
\newblock
{\BBOQ}\APACrefatitle {{Scoring rules for continuous probability
  distributions}} {{Scoring rules for continuous probability
  distributions}}.{\BBCQ}
\newblock
\APACjournalVolNumPages{{Management Science}}{22}{10}{1087--1096}.
\PrintBackRefs{\CurrentBib}

\bibitem [\protect \citeauthoryear {%
McKinnon%
, Rhines%
, Tingley%
\BCBL {}\ \BBA {} Huybers%
}{%
McKinnon%
\ \protect \BOthers {.}}{%
{\protect \APACyear {2016}}%
}]{%
mckinnon2016changing}
\APACinsertmetastar {%
mckinnon2016changing}%
\begin{APACrefauthors}%
McKinnon, K\BPBI A.%
, Rhines, A.%
, Tingley, M\BPBI P.%
\BCBL {}\ \BBA {} Huybers, P.%
\end{APACrefauthors}%
\unskip\
\newblock
\APACrefYearMonthDay{2016}{}{}.
\newblock
{\BBOQ}\APACrefatitle {{The changing shape of Northern Hemisphere summer
  temperature distributions}} {{The changing shape of Northern Hemisphere
  summer temperature distributions}}.{\BBCQ}
\newblock
\APACjournalVolNumPages{{Journal of Geophysical Research:
  Atmospheres}}{121}{15}{8849--8868}.
\PrintBackRefs{\CurrentBib}

\bibitem [\protect \citeauthoryear {%
McLaughlin%
\ \BBA {} Majda%
}{%
McLaughlin%
\ \BBA {} Majda%
}{%
{\protect \APACyear {1996}}%
}]{%
mclaughlin1996explicit}
\APACinsertmetastar {%
mclaughlin1996explicit}%
\begin{APACrefauthors}%
McLaughlin, R\BPBI M.%
\BCBT {}\ \BBA {} Majda, A\BPBI J.%
\end{APACrefauthors}%
\unskip\
\newblock
\APACrefYearMonthDay{1996}{}{}.
\newblock
{\BBOQ}\APACrefatitle {{An explicit example with non-Gaussian probability
  distribution for nontrivial scalar mean and fluctuation}} {{An explicit
  example with non-Gaussian probability distribution for nontrivial scalar mean
  and fluctuation}}.{\BBCQ}
\newblock
\APACjournalVolNumPages{Physics of Fluids}{8}{2}{536--547}.
\PrintBackRefs{\CurrentBib}

\bibitem [\protect \citeauthoryear {%
Newman%
\ \BBA {} Noy%
}{%
Newman%
\ \BBA {} Noy%
}{%
{\protect \APACyear {2023}}%
}]{%
newman2023global}
\APACinsertmetastar {%
newman2023global}%
\begin{APACrefauthors}%
Newman, R.%
\BCBT {}\ \BBA {} Noy, I.%
\end{APACrefauthors}%
\unskip\
\newblock
\APACrefYearMonthDay{2023}{}{}.
\newblock
{\BBOQ}\APACrefatitle {{The global costs of extreme weather that are
  attributable to climate change}} {{The global costs of extreme weather that
  are attributable to climate change}}.{\BBCQ}
\newblock
\APACjournalVolNumPages{{Nature Communications}}{14}{1}{6103}.
\PrintBackRefs{\CurrentBib}

\bibitem [\protect \citeauthoryear {%
Nix%
\ \BBA {} Weigend%
}{%
Nix%
\ \BBA {} Weigend%
}{%
{\protect \APACyear {1994}}%
}]{%
nix1994estimating}
\APACinsertmetastar {%
nix1994estimating}%
\begin{APACrefauthors}%
Nix, D\BPBI A.%
\BCBT {}\ \BBA {} Weigend, A\BPBI S.%
\end{APACrefauthors}%
\unskip\
\newblock
\APACrefYearMonthDay{1994}{}{}.
\newblock
{\BBOQ}\APACrefatitle {{Estimating the mean and variance of the target
  probability distribution}} {{Estimating the mean and variance of the target
  probability distribution}}.{\BBCQ}
\newblock
\BIn{} \APACrefbtitle {{Proceedings of 1994 IEEE International Conference on
  Neural Networks (ICNN'94)}} {{Proceedings of 1994 IEEE International
  Conference on Neural Networks (ICNN'94)}}\ (\BVOL~1, \BPGS\ 55--60).
\PrintBackRefs{\CurrentBib}

\bibitem [\protect \citeauthoryear {%
{NOAA National Centers of Environmental Information}%
}{%
{NOAA National Centers of Environmental Information}%
}{%
{\protect \APACyear {1999}}%
}]{%
noaa1999gsod}
\APACinsertmetastar {%
noaa1999gsod}%
\begin{APACrefauthors}%
{NOAA National Centers of Environmental Information}.%
\end{APACrefauthors}%
\unskip\
\newblock
\APACrefYearMonthDay{1999}{}{}.
\newblock
\APACrefbtitle {{Global Surface Summary of the Day (GSOD), 1.0}.} {{Global
  Surface Summary of the Day (GSOD), 1.0}.}
\newblock
\APAChowpublished
  {\url{https://www.ncei.noaa.gov/data/global-summary-of-the-day/}}.
\newblock
\APACrefnote{Accessed 2021-07-22}
\PrintBackRefs{\CurrentBib}

\bibitem [\protect \citeauthoryear {%
Padilla%
, Tansey%
\BCBL {}\ \BBA {} Chen%
}{%
Padilla%
\ \protect \BOthers {.}}{%
{\protect \APACyear {2022}}%
}]{%
padilla2022quantile}
\APACinsertmetastar {%
padilla2022quantile}%
\begin{APACrefauthors}%
Padilla, O\BPBI H\BPBI M.%
, Tansey, W.%
\BCBL {}\ \BBA {} Chen, Y.%
\end{APACrefauthors}%
\unskip\
\newblock
\APACrefYearMonthDay{2022}{}{}.
\newblock
{\BBOQ}\APACrefatitle {{Quantile regression with ReLU Networks: Estimators and
  minimax rates}} {{Quantile regression with ReLU Networks: Estimators and
  minimax rates}}.{\BBCQ}
\newblock
\APACjournalVolNumPages{{Journal of Machine Learning
  Research}}{23}{247}{1--42}.
\newblock
\APACrefnote{\url{http://jmlr.org/papers/v23/21-0309.html}}
\PrintBackRefs{\CurrentBib}

\bibitem [\protect \citeauthoryear {%
Papacharalampous%
, Tyralis%
, Doulamis%
\BCBL {}\ \BBA {} Doulamis%
}{%
Papacharalampous%
\ \protect \BOthers {.}}{%
{\protect \APACyear {2025}}%
}]{%
papacharalampous2025ensemble}
\APACinsertmetastar {%
papacharalampous2025ensemble}%
\begin{APACrefauthors}%
Papacharalampous, G.%
, Tyralis, H.%
, Doulamis, N.%
\BCBL {}\ \BBA {} Doulamis, A.%
\end{APACrefauthors}%
\unskip\
\newblock
\APACrefYearMonthDay{2025}{}{}.
\newblock
{\BBOQ}\APACrefatitle {{Ensemble learning for uncertainty estimation with
  application to the correction of satellite precipitation products}}
  {{Ensemble learning for uncertainty estimation with application to the
  correction of satellite precipitation products}}.{\BBCQ}
\newblock
\APACjournalVolNumPages{Machine Learning: Earth}{1}{1}{015004}.
\PrintBackRefs{\CurrentBib}

\bibitem [\protect \citeauthoryear {%
Penland%
}{%
Penland%
}{%
{\protect \APACyear {1989}}%
}]{%
penland1989random}
\APACinsertmetastar {%
penland1989random}%
\begin{APACrefauthors}%
Penland, C.%
\end{APACrefauthors}%
\unskip\
\newblock
\APACrefYearMonthDay{1989}{}{}.
\newblock
{\BBOQ}\APACrefatitle {{Random Forcing and Forecasting Using Principal
  Oscillation Pattern Analysis}} {{Random Forcing and Forecasting Using
  Principal Oscillation Pattern Analysis}}.{\BBCQ}
\newblock
\APACjournalVolNumPages{{Monthly Weather Review}}{117}{10}{2165--2185}.
\PrintBackRefs{\CurrentBib}

\bibitem [\protect \citeauthoryear {%
Penland%
\ \BBA {} Sardeshmukh%
}{%
Penland%
\ \BBA {} Sardeshmukh%
}{%
{\protect \APACyear {1995}}%
}]{%
penland1995optimal}
\APACinsertmetastar {%
penland1995optimal}%
\begin{APACrefauthors}%
Penland, C.%
\BCBT {}\ \BBA {} Sardeshmukh, P\BPBI D.%
\end{APACrefauthors}%
\unskip\
\newblock
\APACrefYearMonthDay{1995}{}{}.
\newblock
{\BBOQ}\APACrefatitle {{The Optimal Growth of Tropical Sea Surface Temperature
  Anomalies}} {{The Optimal Growth of Tropical Sea Surface Temperature
  Anomalies}}.{\BBCQ}
\newblock
\APACjournalVolNumPages{{Journal of Climate}}{8}{8}{1999--2024}.
\PrintBackRefs{\CurrentBib}

\bibitem [\protect \citeauthoryear {%
Perezhogin%
, Zanna%
\BCBL {}\ \BBA {} Fernandez-Granda%
}{%
Perezhogin%
\ \protect \BOthers {.}}{%
{\protect \APACyear {2023}}%
}]{%
perezhogin2023generative}
\APACinsertmetastar {%
perezhogin2023generative}%
\begin{APACrefauthors}%
Perezhogin, P.%
, Zanna, L.%
\BCBL {}\ \BBA {} Fernandez-Granda, C.%
\end{APACrefauthors}%
\unskip\
\newblock
\APACrefYearMonthDay{2023}{}{}.
\newblock
{\BBOQ}\APACrefatitle {{Generative Data-Driven Approaches for Stochastic
  Subgrid Parameterizations in an Idealized Ocean Model}} {{Generative
  Data-Driven Approaches for Stochastic Subgrid Parameterizations in an
  Idealized Ocean Model}}.{\BBCQ}
\newblock
\APACjournalVolNumPages{Journal of Advances in Modeling Earth
  Systems}{15}{10}{e2023MS003681}.
\PrintBackRefs{\CurrentBib}

\bibitem [\protect \citeauthoryear {%
Proistosescu%
, Rhines%
\BCBL {}\ \BBA {} Huybers%
}{%
Proistosescu%
\ \protect \BOthers {.}}{%
{\protect \APACyear {2016}}%
}]{%
proistosescu2016identification}
\APACinsertmetastar {%
proistosescu2016identification}%
\begin{APACrefauthors}%
Proistosescu, C.%
, Rhines, A.%
\BCBL {}\ \BBA {} Huybers, P.%
\end{APACrefauthors}%
\unskip\
\newblock
\APACrefYearMonthDay{2016}{}{}.
\newblock
{\BBOQ}\APACrefatitle {{Identification and interpretation of nonnormality in
  atmospheric time series}} {{Identification and interpretation of nonnormality
  in atmospheric time series}}.{\BBCQ}
\newblock
\APACjournalVolNumPages{Geophysical Research Letters}{43}{10}{5425--5434}.
\PrintBackRefs{\CurrentBib}

\bibitem [\protect \citeauthoryear {%
Scheuerer%
, Switanek%
, Worsnop%
\BCBL {}\ \BBA {} Hamill%
}{%
Scheuerer%
\ \protect \BOthers {.}}{%
{\protect \APACyear {2020}}%
}]{%
scheuerer2020using}
\APACinsertmetastar {%
scheuerer2020using}%
\begin{APACrefauthors}%
Scheuerer, M.%
, Switanek, M\BPBI B.%
, Worsnop, R\BPBI P.%
\BCBL {}\ \BBA {} Hamill, T\BPBI M.%
\end{APACrefauthors}%
\unskip\
\newblock
\APACrefYearMonthDay{2020}{}{}.
\newblock
{\BBOQ}\APACrefatitle {{Using Artificial Neural Networks for Generating
  Probabilistic Subseasonal Precipitation Forecasts over California}} {{Using
  Artificial Neural Networks for Generating Probabilistic Subseasonal
  Precipitation Forecasts over California}}.{\BBCQ}
\newblock
\APACjournalVolNumPages{Monthly Weather Review}{148}{8}{3489--3506}.
\PrintBackRefs{\CurrentBib}

\bibitem [\protect \citeauthoryear {%
Schlick%
}{%
Schlick%
}{%
{\protect \APACyear {2010}}%
}]{%
schlick2010molecular}
\APACinsertmetastar {%
schlick2010molecular}%
\begin{APACrefauthors}%
Schlick, T.%
\end{APACrefauthors}%
\unskip\
\newblock
\APACrefYear{2010}.
\newblock
\APACrefbtitle {{Molecular Modeling and Simulation: An Interdisciplinary
  Guide}} {{Molecular Modeling and Simulation: An Interdisciplinary Guide}}\
  (\BVOL~2).
\newblock
\APACaddressPublisher{}{Springer}.
\PrintBackRefs{\CurrentBib}

\bibitem [\protect \citeauthoryear {%
Schreck%
\ \protect \BOthers {.}}{%
Schreck%
\ \protect \BOthers {.}}{%
{\protect \APACyear {2024}}%
}]{%
schreck2024evidential}
\APACinsertmetastar {%
schreck2024evidential}%
\begin{APACrefauthors}%
Schreck, J\BPBI S.%
, Gagne, D\BPBI J.%
, Becker, C.%
, Chapman, W\BPBI E.%
, Elmore, K.%
, Fan, D.%
\BDBL {}others%
\end{APACrefauthors}%
\unskip\
\newblock
\APACrefYearMonthDay{2024}{}{}.
\newblock
{\BBOQ}\APACrefatitle {{Evidential Deep Learning: Enhancing Predictive
  Uncertainty Estimation for Earth System Science Applications}} {{Evidential
  Deep Learning: Enhancing Predictive Uncertainty Estimation for Earth System
  Science Applications}}.{\BBCQ}
\newblock
\APACjournalVolNumPages{Artificial Intelligence for the Earth
  Systems}{3}{4}{230093}.
\PrintBackRefs{\CurrentBib}

\bibitem [\protect \citeauthoryear {%
Seitzer%
, Tavakoli%
, Antic%
\BCBL {}\ \BBA {} Martius%
}{%
Seitzer%
\ \protect \BOthers {.}}{%
{\protect \APACyear {2022}}%
}]{%
seitzer2022on}
\APACinsertmetastar {%
seitzer2022on}%
\begin{APACrefauthors}%
Seitzer, M.%
, Tavakoli, A.%
, Antic, D.%
\BCBL {}\ \BBA {} Martius, G.%
\end{APACrefauthors}%
\unskip\
\newblock
\APACrefYearMonthDay{2022}{}{}.
\newblock
{\BBOQ}\APACrefatitle {{On the Pitfalls of Heteroscedastic Uncertainty
  Estimation with Probabilistic Neural Networks}} {{On the Pitfalls of
  Heteroscedastic Uncertainty Estimation with Probabilistic Neural
  Networks}}.{\BBCQ}
\newblock
\BIn{} \APACrefbtitle {International Conference on Learning Representations.}
  {International conference on learning representations.}
\newblock
\APACrefnote{\url{https://openreview.net/forum?id=aPOpXlnV1T}}
\PrintBackRefs{\CurrentBib}

\bibitem [\protect \citeauthoryear {%
Sluijterman%
, Cator%
\BCBL {}\ \BBA {} Heskes%
}{%
Sluijterman%
\ \protect \BOthers {.}}{%
{\protect \APACyear {2024}}%
}]{%
sluijterman2024optimal}
\APACinsertmetastar {%
sluijterman2024optimal}%
\begin{APACrefauthors}%
Sluijterman, L.%
, Cator, E.%
\BCBL {}\ \BBA {} Heskes, T.%
\end{APACrefauthors}%
\unskip\
\newblock
\APACrefYearMonthDay{2024}{}{}.
\newblock
{\BBOQ}\APACrefatitle {{Optimal training of Mean Variance Estimation neural
  networks}} {{Optimal training of Mean Variance Estimation neural
  networks}}.{\BBCQ}
\newblock
\APACjournalVolNumPages{{Neurocomputing}}{597}{}{127929}.
\PrintBackRefs{\CurrentBib}

\bibitem [\protect \citeauthoryear {%
A.~Smith%
, Lott%
\BCBL {}\ \BBA {} Vose%
}{%
A.~Smith%
\ \protect \BOthers {.}}{%
{\protect \APACyear {2011}}%
}]{%
smith2011integrated}
\APACinsertmetastar {%
smith2011integrated}%
\begin{APACrefauthors}%
Smith, A.%
, Lott, N.%
\BCBL {}\ \BBA {} Vose, R.%
\end{APACrefauthors}%
\unskip\
\newblock
\APACrefYearMonthDay{2011}{}{}.
\newblock
{\BBOQ}\APACrefatitle {{The Integrated Surface Database: Recent Developments
  and Partnerships}} {{The Integrated Surface Database: Recent Developments and
  Partnerships}}.{\BBCQ}
\newblock
\APACjournalVolNumPages{Bulletin of the American Meteorological
  Society}{92}{6}{704--708}.
\PrintBackRefs{\CurrentBib}

\bibitem [\protect \citeauthoryear {%
L\BPBI N.~Smith%
}{%
L\BPBI N.~Smith%
}{%
{\protect \APACyear {2018}}%
}]{%
smith2018disciplined}
\APACinsertmetastar {%
smith2018disciplined}%
\begin{APACrefauthors}%
Smith, L\BPBI N.%
\end{APACrefauthors}%
\unskip\
\newblock
\APACrefYearMonthDay{2018}{}{}.
\newblock
\APACrefbtitle {{A disciplined approach to neural network hyper-parameters:
  Part 1 -- learning rate, batch size, momentum, and weight decay}.} {{A
  disciplined approach to neural network hyper-parameters: Part 1 -- learning
  rate, batch size, momentum, and weight decay}.}
\newblock
\begin{APACrefURL} \url{https://arxiv.org/abs/1803.09820} \end{APACrefURL}
\PrintBackRefs{\CurrentBib}

\bibitem [\protect \citeauthoryear {%
Sura%
\ \BBA {} Hannachi%
}{%
Sura%
\ \BBA {} Hannachi%
}{%
{\protect \APACyear {2015}}%
}]{%
sura2015perspectives}
\APACinsertmetastar {%
sura2015perspectives}%
\begin{APACrefauthors}%
Sura, P.%
\BCBT {}\ \BBA {} Hannachi, A.%
\end{APACrefauthors}%
\unskip\
\newblock
\APACrefYearMonthDay{2015}{}{}.
\newblock
{\BBOQ}\APACrefatitle {{Perspectives of non-Gaussianity in atmospheric synoptic
  and low-frequency variability}} {{Perspectives of non-Gaussianity in
  atmospheric synoptic and low-frequency variability}}.{\BBCQ}
\newblock
\APACjournalVolNumPages{Journal of Climate}{28}{13}{5091--5114}.
\PrintBackRefs{\CurrentBib}

\bibitem [\protect \citeauthoryear {%
Taggart%
}{%
Taggart%
}{%
{\protect \APACyear {2023}}%
}]{%
taggart2023estimation}
\APACinsertmetastar {%
taggart2023estimation}%
\begin{APACrefauthors}%
Taggart, R.%
\end{APACrefauthors}%
\unskip\
\newblock
\APACrefYear{2023}.
\newblock
\APACrefbtitle {{Estimation of CRPS for precipitation forecasts using weighted
  sums of quantile scores and Brier scores}} {{Estimation of CRPS for
  precipitation forecasts using weighted sums of quantile scores and Brier
  scores}}.
\newblock
\APACaddressPublisher{}{Bureau of Meteorology}.
\PrintBackRefs{\CurrentBib}

\bibitem [\protect \citeauthoryear {%
Tambwekar%
, Maiya%
, Dhavala%
\BCBL {}\ \BBA {} Saha%
}{%
Tambwekar%
\ \protect \BOthers {.}}{%
{\protect \APACyear {2021}}%
}]{%
tambwekar2021estimation}
\APACinsertmetastar {%
tambwekar2021estimation}%
\begin{APACrefauthors}%
Tambwekar, A.%
, Maiya, A.%
, Dhavala, S.%
\BCBL {}\ \BBA {} Saha, S.%
\end{APACrefauthors}%
\unskip\
\newblock
\APACrefYearMonthDay{2021}{}{}.
\newblock
{\BBOQ}\APACrefatitle {{Estimation and applications of quantiles in deep binary
  classification}} {{Estimation and applications of quantiles in deep binary
  classification}}.{\BBCQ}
\newblock
\APACjournalVolNumPages{{IEEE Transactions on Artificial
  Intelligence}}{3}{2}{275--286}.
\PrintBackRefs{\CurrentBib}

\bibitem [\protect \citeauthoryear {%
Taylor%
}{%
Taylor%
}{%
{\protect \APACyear {2000}}%
}]{%
taylor2000quantile}
\APACinsertmetastar {%
taylor2000quantile}%
\begin{APACrefauthors}%
Taylor, J\BPBI W.%
\end{APACrefauthors}%
\unskip\
\newblock
\APACrefYearMonthDay{2000}{}{}.
\newblock
{\BBOQ}\APACrefatitle {{A quantile regression neural network approach to
  estimating the conditional density of multiperiod returns}} {{A quantile
  regression neural network approach to estimating the conditional density of
  multiperiod returns}}.{\BBCQ}
\newblock
\APACjournalVolNumPages{{Journal of Forecasting}}{19}{4}{299--311}.
\PrintBackRefs{\CurrentBib}

\bibitem [\protect \citeauthoryear {%
Vitart%
\ \protect \BOthers {.}}{%
Vitart%
\ \protect \BOthers {.}}{%
{\protect \APACyear {2017}}%
}]{%
vitart2017subseasonal}
\APACinsertmetastar {%
vitart2017subseasonal}%
\begin{APACrefauthors}%
Vitart, F.%
, Ardilouze, C.%
, Bonet, A.%
, Brookshaw, A.%
, Chen, M.%
, Codorean, C.%
\BDBL {}others%
\end{APACrefauthors}%
\unskip\
\newblock
\APACrefYearMonthDay{2017}{}{}.
\newblock
{\BBOQ}\APACrefatitle {{The Subseasonal to Seasonal (S2S) Prediction Project
  Database}} {{The Subseasonal to Seasonal (S2S) Prediction Project
  Database}}.{\BBCQ}
\newblock
\APACjournalVolNumPages{{Bulletin of the American Meteorological
  Society}}{98}{1}{163--173}.
\PrintBackRefs{\CurrentBib}

\bibitem [\protect \citeauthoryear {%
Wasserman%
}{%
Wasserman%
}{%
{\protect \APACyear {2004}}%
}]{%
wasserman2004all}
\APACinsertmetastar {%
wasserman2004all}%
\begin{APACrefauthors}%
Wasserman, L.%
\end{APACrefauthors}%
\unskip\
\newblock
\APACrefYear{2004}.
\newblock
\APACrefbtitle {{Parametric Inference}} {{Parametric Inference}}\ (\BVOL~26).
\newblock
\APACaddressPublisher{}{Springer}.
\PrintBackRefs{\CurrentBib}

\bibitem [\protect \citeauthoryear {%
White%
}{%
White%
}{%
{\protect \APACyear {1992}}%
}]{%
white1992nonparametric}
\APACinsertmetastar {%
white1992nonparametric}%
\begin{APACrefauthors}%
White, H.%
\end{APACrefauthors}%
\unskip\
\newblock
\APACrefYearMonthDay{1992}{}{}.
\newblock
{\BBOQ}\APACrefatitle {{Nonparametric estimation of conditional quantiles using
  neural networks}} {{Nonparametric estimation of conditional quantiles using
  neural networks}}.{\BBCQ}
\newblock
\BIn{} \APACrefbtitle {Computing Science and Statistics: Statistics of Many
  Parameters: Curves, Images, Spatial Models} {Computing science and
  statistics: Statistics of many parameters: Curves, images, spatial models}\
  (\BPGS\ 190--199).
\newblock
\APACaddressPublisher{}{Springer}.
\PrintBackRefs{\CurrentBib}

\bibitem [\protect \citeauthoryear {%
Wu%
\ \protect \BOthers {.}}{%
Wu%
\ \protect \BOthers {.}}{%
{\protect \APACyear {2025}}%
}]{%
wu2025data}
\APACinsertmetastar {%
wu2025data}%
\begin{APACrefauthors}%
Wu, J.%
, Perezhogin, P.%
, Gagne, D\BPBI J.%
, Reichl, B.%
, Subramanian, A\BPBI C.%
, Thompson, E.%
\BCBL {}\ \BBA {} Zanna, L.%
\end{APACrefauthors}%
\unskip\
\newblock
\APACrefYearMonthDay{2025}{}{}.
\newblock
\APACrefbtitle {{Data-Driven Probabilistic Air-Sea Flux Parameterization}.}
  {{Data-Driven Probabilistic Air-Sea Flux Parameterization}.}
\newblock
\APACrefnote{\url{https://arxiv.org/abs/2503.03990}}
\PrintBackRefs{\CurrentBib}

\bibitem [\protect \citeauthoryear {%
Xu%
, Deng%
, Jiang%
, Sun%
\BCBL {}\ \BBA {} Huang%
}{%
Xu%
\ \protect \BOthers {.}}{%
{\protect \APACyear {2017}}%
}]{%
xu2017composite}
\APACinsertmetastar {%
xu2017composite}%
\begin{APACrefauthors}%
Xu, Q.%
, Deng, K.%
, Jiang, C.%
, Sun, F.%
\BCBL {}\ \BBA {} Huang, X.%
\end{APACrefauthors}%
\unskip\
\newblock
\APACrefYearMonthDay{2017}{}{}.
\newblock
{\BBOQ}\APACrefatitle {{Composite quantile regression neural network with
  applications}} {{Composite quantile regression neural network with
  applications}}.{\BBCQ}
\newblock
\APACjournalVolNumPages{{Expert Systems with Applications}}{76}{}{129--139}.
\PrintBackRefs{\CurrentBib}

\bibitem [\protect \citeauthoryear {%
W.~Zhang%
, Quan%
\BCBL {}\ \BBA {} Srinivasan%
}{%
W.~Zhang%
\ \protect \BOthers {.}}{%
{\protect \APACyear {2019}}%
}]{%
zhang2019improved}
\APACinsertmetastar {%
zhang2019improved}%
\begin{APACrefauthors}%
Zhang, W.%
, Quan, H.%
\BCBL {}\ \BBA {} Srinivasan, D.%
\end{APACrefauthors}%
\unskip\
\newblock
\APACrefYearMonthDay{2019}{}{}.
\newblock
{\BBOQ}\APACrefatitle {{An Improved Quantile Regression Neural Network for
  Probabilistic Load Forecasting}} {{An Improved Quantile Regression Neural
  Network for Probabilistic Load Forecasting}}.{\BBCQ}
\newblock
\APACjournalVolNumPages{{IEEE Transactions on Smart Grid}}{10}{4}{4425--4434}.
\newblock
\begin{APACrefDOI} \doi{10.1109/TSG.2018.2859749} \end{APACrefDOI}
\PrintBackRefs{\CurrentBib}

\bibitem [\protect \citeauthoryear {%
X.~Zhang%
, Yuan%
, Wang%
\BCBL {}\ \BBA {} Song%
}{%
X.~Zhang%
\ \protect \BOthers {.}}{%
{\protect \APACyear {2025}}%
}]{%
zhang2025monotone}
\APACinsertmetastar {%
zhang2025monotone}%
\begin{APACrefauthors}%
Zhang, X.%
, Yuan, X.%
, Wang, C.%
\BCBL {}\ \BBA {} Song, X.%
\end{APACrefauthors}%
\unskip\
\newblock
\APACrefYearMonthDay{2025}{}{}.
\newblock
{\BBOQ}\APACrefatitle {{Monotone composite quantile regression neural network
  for censored data with a cure fraction}} {{Monotone composite quantile
  regression neural network for censored data with a cure fraction}}.{\BBCQ}
\newblock
\APACjournalVolNumPages{Computational Statistics \& Data Analysis}{}{}{108201}.
\PrintBackRefs{\CurrentBib}

\bibitem [\protect \citeauthoryear {%
Zou%
\ \BBA {} Yuan%
}{%
Zou%
\ \BBA {} Yuan%
}{%
{\protect \APACyear {2008}}%
}]{%
zou2008composite}
\APACinsertmetastar {%
zou2008composite}%
\begin{APACrefauthors}%
Zou, H.%
\BCBT {}\ \BBA {} Yuan, M.%
\end{APACrefauthors}%
\unskip\
\newblock
\APACrefYearMonthDay{2008}{}{}.
\newblock
{\BBOQ}\APACrefatitle {{Composite Quantile Regression and the Oracle Model
  Selection Theory}} {{Composite Quantile Regression and the Oracle Model
  Selection Theory}}.{\BBCQ}
\newblock
\APACjournalVolNumPages{{Annals of Statistics}}{36}{3}{1108--1126}.
\newblock
\begin{APACrefDOI} \doi{10.1214/07-AOS50} \end{APACrefDOI}
\PrintBackRefs{\CurrentBib}

\end{thebibliography}

%
%
%
%
%

\end{document}